\documentclass{article} % For LaTeX2e
\usepackage{iclr2026_conference,times}

\usepackage{amsmath,amsfonts,bm}

\def\eqref#1{equation~\ref{#1}}
\def\1{\bm{1}}

\DeclareMathAlphabet{\mathsfit}{\encodingdefault}{\sfdefault}{m}{sl}
\SetMathAlphabet{\mathsfit}{bold}{\encodingdefault}{\sfdefault}{bx}{n}

\usepackage{hyperref}
\usepackage{url}
\usepackage{multirow}
\usepackage{array}
\usepackage{booktabs} 
\usepackage{graphicx} 
\usepackage{subcaption} 
\usepackage{wrapfig}

\title{Toward Complex-Valued Neural Networks for Waveform Generation}

\author{Hyung-Seok~Oh, Deok-Hyeon~Cho, Seung-Bin~Kim \& Seong-Whan~Lee \thanks{Corresponding author} \\
Department of Artificial Intelligence\\
Korea University\\
Seoul, Republic of Korea \\
\texttt{\{hs\_oh, dh\_cho, sb-kim, sw.lee\}@korea.ac.kr}
}

\iclrfinalcopy % Uncomment for camera-ready version, but NOT for submission.
\begin{document}

\maketitle

\begin{abstract}
Neural vocoders have recently advanced waveform generation, yielding natural and expressive audio.
Among these approaches, iSTFT-based vocoders have recently gained attention. 
They predict a complex-valued spectrogram and then synthesize the waveform via iSTFT, thereby avoiding learned upsampling stages that can increase computational cost. 
However, current approaches use real-valued networks that process the real and imaginary parts independently.
This separation limits their ability to capture the inherent structure of complex spectrograms.
We present ComVo, a \textbf{Com}plex-valued neural \textbf{Vo}coder whose generator and discriminator use native complex arithmetic.
This enables an adversarial training framework that provides structured feedback in complex-valued representations.
To guide phase transformations in a structured manner, we introduce phase quantization, which discretizes phase values and regularizes the training process.
Finally, we propose a block-matrix computation scheme to improve training efficiency by reducing redundant operations. 
Experiments demonstrate that ComVo achieves higher synthesis quality than comparable real-valued baselines, and that its block-matrix scheme reduces training time by 25\%.
Audio samples and code are available at \url{https://hs-oh-prml.github.io/ComVo/}.
\end{abstract}

\section{Introduction}
Deep learning-based vocoders have significantly advanced speech synthesis, producing more natural and expressive synthetic speech.
Recent developments include models based on generative adversarial networks (GANs) \citep{NEURIPS2019_6804c9bc, 9053795, NEURIPS2020_c5d73680, lee2023bigvgan}, normalizing flow-based models \citep{pmlr-v80-oord18a, pmlr-v119-ping20a, NEURIPS2020_a1c3ae6c}, and diffusion-based models \citep{kong2021diffwave, lee2022priorgrad, chen2021wavegrad, lee2025periodwave}.
Although these approaches achieve high-fidelity speech generation, some neural vocoders still rely on sequential sample prediction or learned upsampling, thereby increasing model complexity and inference latency.

An alternative is to synthesize speech in the spectral domain using the inverse short-time Fourier transform (iSTFT).
Operating directly on complex spectrograms \citep{8553396, neekhara19_interspeech, NEURIPS2020_9873eaad, 9746713, kaneko23_interspeech, siuzdak2024vocos, yoneyama2024wavehax, liu2025rfwave} avoids the need for sample-by-sample generation and learned upsampling.
To our knowledge, current iSTFT-based vocoders rely on real-valued neural networks (RVNNs) that process real and imaginary parts as separate channels.
This separation limits their ability to model the coupling between these components.

Complex-valued neural networks (CVNNs) extend standard neural networks to the complex domain by allowing both inputs and parameters to be complex-valued.
Operating entirely in the complex domain enables these models to capture the intrinsic dependencies between the real and imaginary components.
CVNNs have been applied in domains such as radar signal classification \citep{9763903}, MRI reconstruction \citep{Vasudeva_2022_WACV}, and wireless communication \citep{9766131}, where measurements carry both magnitude and phase information and naturally form complex-valued data.
In speech processing, CVNNs have been explored for tasks including speech enhancement \citep{10637717, mamun23_interspeech}, speech recognition \citep{8659610}, and even statistical parametric speech synthesis \citep{7472755}.
These studies demonstrate the potential of CVNNs to better capture spectral structure.

Although some recent vocoders produce complex spectrograms, they still use real-valued networks that handle each spectrogram channel independently.  
CVNNs, by jointly processing complex coefficients, could overcome this limitation.
By treating each spectrogram coefficient as a unified complex entity, CVNN-based models can capture cross-component interactions that real-valued models miss. 
Motivated by this, we adopt CVNNs to better capture structure in the complex domain, yielding higher-quality synthesis.

In this work, we propose ComVo, a \textbf{Com}plex-valued neural \textbf{Vo}coder that performs iSTFT-based waveform generation entirely in the complex domain with a GAN-based architecture.
The generator uses CVNN layers to jointly model the real and imaginary components of spectrograms, thereby better capturing their algebraic structure.
We then design a complex multi-resolution discriminator (cMRD) that operates directly on complex spectrograms.
Together, these components form a complex-domain adversarial training framework in which both the generator and discriminator operate on complex-valued representations.
This design allows feedback that respects the structure of the complex domain.
Inspired by recent studies on complex activation functions \citep{Vasudeva_2022_WACV}, we introduce phase quantization, a nonlinear transformation that discretizes phase angles to serve as an inductive bias for stable learning. 
Finally, to reduce redundant computations in complex-valued operations, we develop a block-matrix computation scheme that improves overall training efficiency.

\begin{itemize}
  \item {\bf CVNN-based architecture with complex adversarial training:} 
  We introduce ComVo, which, to our knowledge, is the first iSTFT-based vocoder
to employ complex-valued neural networks in both its generator and discriminator. 
  We design the discriminator losses in the complex domain, thus establishing an adversarial framework that operates on complex-valued representations.

  \item {\bf Structured nonlinear transformation:}
  We propose phase quantization, a tailored nonlinear operation that discretizes phase angles and serves as an inductive bias.
  
  \item {\bf Block-matrix computation scheme:} 
  We present an efficient implementation that fuses the four real-valued multiplications required for each complex operation into a single block-matrix multiplication, reducing training time by ~25\%.
  \item {\bf Improved synthesis performance:} 
  ComVo outperforms real-valued vocoders, as demonstrated in our experiments.
\end{itemize}

\section{Related works}
\subsection{Complex-valued Neural Networks}         

CVNNs represent inputs, activations, and weights directly as complex numbers.
They have been applied in a range of domains where signals are naturally expressed in the complex field, including radar classification \citep{9763903}, MRI reconstruction \citep{Vasudeva_2022_WACV}, wireless communication \citep{9766131}, and audio analysis \citep{sarroff2018complex}.
Several studies report that CVNNs can exhibit favorable learning behavior or approximation properties compared to real-valued networks in various settings \citep{9413814, VOIGTLAENDER202333, NEURIPS2023_05b69cc4}.
This prior work suggests that complex-valued modeling can be a viable choice when dealing with data or transformations formulated in the complex domain.

\subsection{iSTFT-based Vocoder}
The short-time Fourier transform (STFT) decomposes a waveform into overlapping frames of complex spectral coefficients. 
The iSTFT reconstructs the time-domain signal using the overlap-add method.
This fully differentiable analysis-synthesis pipeline enables end-to-end training on frame-level spectra while generating sample-level waveforms in a single pass. 
This approach eliminates any explicit upsampling or autoregressive generation, thereby reducing latency. 
Early methods, such as the Griffin-Lim algorithm \citep{1164317}, used iterative phase reconstruction but often yielded suboptimal coherence between magnitude and phase. 
GLA-Grad \citep{10446058} later combined Griffin-Lim with neural diffusion models to improve phase accuracy.

More recent neural iSTFT-based vocoders, such as iSTFTNet \citep{9746713}, iSTFTNet2 \citep{kaneko23_interspeech}, APNet \citep{10128683}, APNet2 \citep{10.1007/978-981-97-0601-3_6}, FreeV \citep{lv24_interspeech}, Vocos \citep{siuzdak2024vocos}, and RFWave \citep{liu2025rfwave}, employ diverse architectural designs for iSTFT-based waveform generation.

In these systems, the STFT-domain coefficients are generated directly for frame-level synthesis, enabling efficient inference without waveform upsampling. 
Our work retains this benefit but additionally focuses on how this representation is modeled within the network. 
For this reason, we use complex-valued layers that operate directly in the complex domain rather than separating each coefficient into real and imaginary channels.

\section{Preliminary Analysis of Real- and Complex-Valued Networks}
Recent work on complex-valued neural networks suggests that operating directly in the complex field can better capture interactions between a variable’s magnitude and phase than relying on real-valued parameterizations that treat the two components independently \citep{9413814, DOU2025108685}.
Motivated by this perspective, we conduct a controlled generative experiment designed to isolate the effect of complex-domain modeling from architectural factors specific to waveform generation.

\begin{figure*}[!t]
\centering
\includegraphics[width=0.9\linewidth]{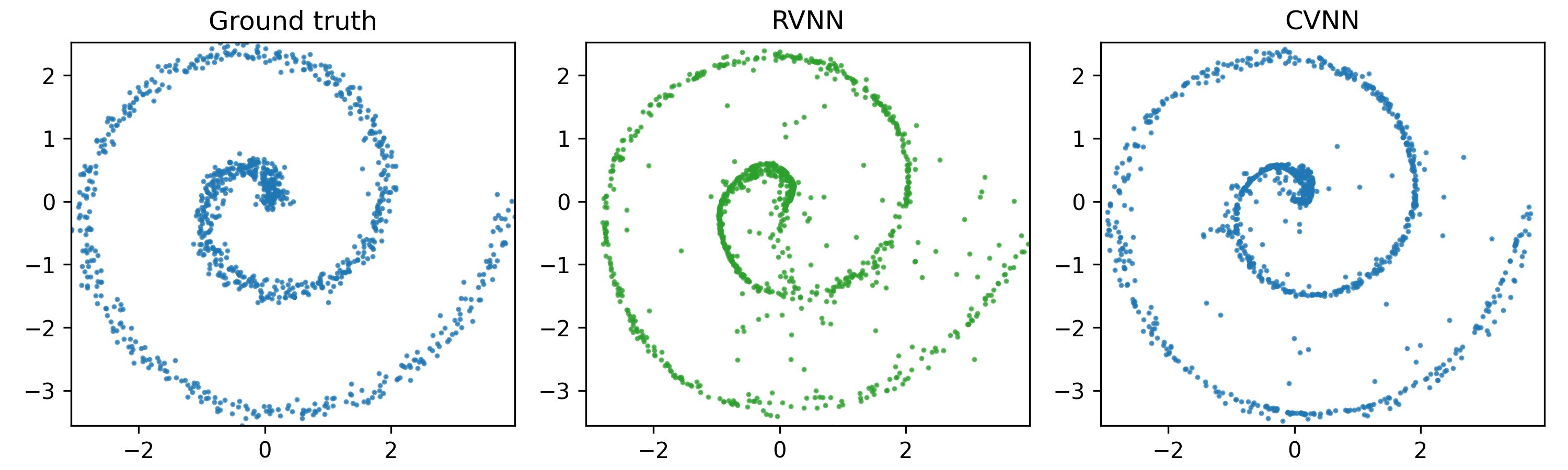}
\caption{Ground-truth distribution compared with samples generated by RVNN and CVNN.}
\label{fig:rvnn_cvnn}
\vspace{-0.3cm}
\end{figure*}

\begin{wraptable}{r}{0.48\linewidth}
\vspace{-0.3cm}
\centering

\caption{ JSD between the generated and ground-truth magnitude (mag.) and phase distributions for RVNN and CVNN.}
\label{tab:rvnn_cvnn_jsd}

\resizebox{\linewidth}{!}{%
\begin{tabular}{l|cc}
\toprule
Model  & JSD (mag.)            & JSD (phase) \\
\midrule
RVNN   & 0.018350 $\pm$ 0.014  & 0.021110 $\pm$ 0.036 \\
CVNN   & 0.006548 $\pm$ 0.003  & 0.003911 $\pm$ 0.002 \\
\bottomrule
\end{tabular}
} % end resizebox
\vspace{-0.2cm}
\end{wraptable}

We train a lightweight MLP-based GAN on a synthetic complex distribution and compare two models: RVNN, which represents complex numbers as two real channels, and CVNN, which processes each coefficient as a single complex entity.
Because the CVNN stores real and imaginary parameters separately, it requires roughly twice the memory for a given layer width; to match memory usage fairly, the RVNN is assigned twice the hidden dimension.

Figure~\ref{fig:rvnn_cvnn} presents sample visualizations across multiple training seeds, and Table~\ref{tab:rvnn_cvnn_jsd} reports the Jensen–Shannon divergence (JSD) between the generated and target magnitude and phase distributions, computed using a kernel density–based estimator.
Both models recover the broad structure of the target distribution, but the CVNN yields samples that adhere more closely to the underlying trajectory and exhibit lower JSD in both magnitude and phase.

These observations provide a simple, controlled example in which modeling directly in the complex domain offers representational advantages when the data possess inherent real–imaginary dependencies.
This motivates our use of CVNNs in the proposed method that follows.
Extended analysis and additional visualizations are included in the Appendix~\ref{sec:analysis_real_complex}.

\section{Method}
We present ComVo, an iSTFT-based GAN vocoder whose generator and discriminator operate entirely in the complex domain, preserving real-imaginary interactions end to end.
The model uses an iSTFT synthesis pipeline with adversarial training objectives. 
We also include a phase quantization layer as an inductive bias and adopt a block-matrix formulation for efficient complex-valued computation.
Figure~\ref{fig:model} provides an overview of the architecture.

\subsection{Generator}
\begin{figure*}[!t]
\centering
\includegraphics[width=1.0\linewidth]{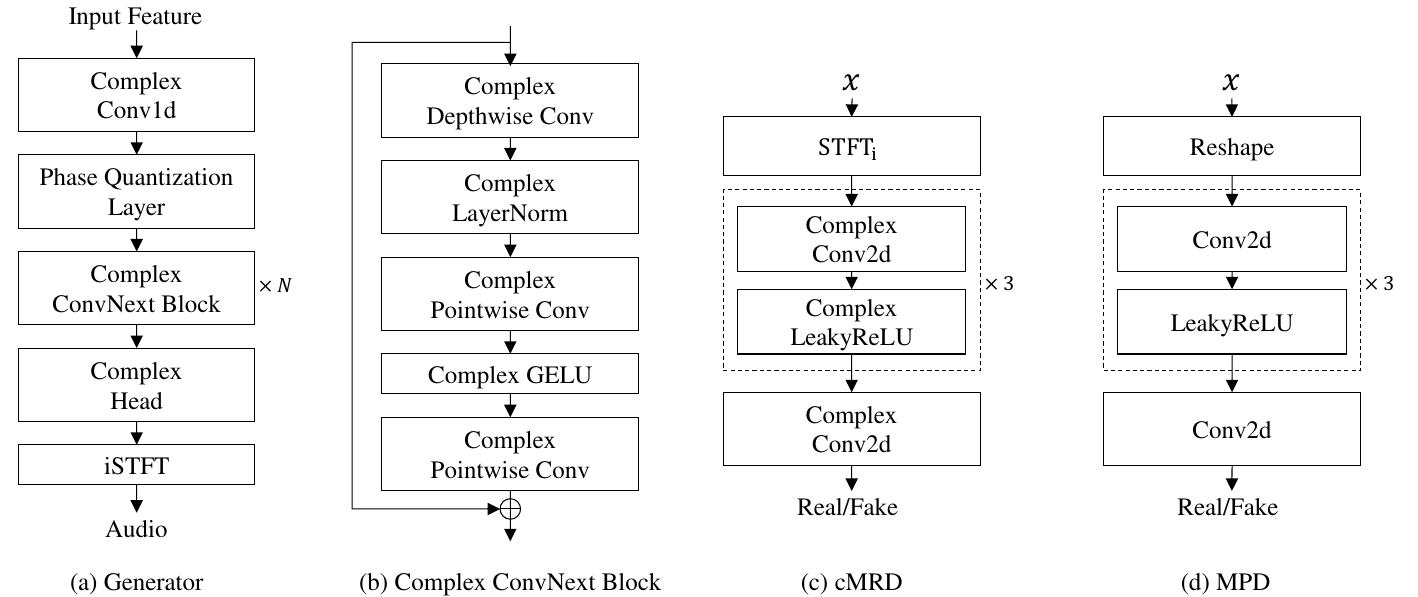}
\caption{Overview of the ComVo architecture.}
\label{fig:model}
\vspace{-0.3cm}
\end{figure*}

Figure~\ref{fig:model}(a) depicts our generator, which is adapted from the Vocos architecture~\citep{siuzdak2024vocos}. 
We chose Vocos as our starting point because it synthesizes via frame-level iSTFT without requiring learned upsampling, features a compact feed-forward structure, and serves as a widely used baseline for comparison.
All convolutions and normalizations in our generator are implemented in the complex domain.
We use a split GELU activation~\citep{hendrycks2016gaussian} to maintain the ConvNeXt-style block layout in the complex setting.
After the initial complex convolution, a phase quantization layer discretizes phase values to stabilize training.
Figure~\ref{fig:model}(b) details the complex ConvNeXt block used at each generator stage.

\subsection{Discriminator}
We propose a complex multi-resolution discriminator (cMRD), as shown in Figure~\ref{fig:model}(c).
Prior work on spectrogram-based discriminators typically used either only magnitude spectra or concatenated the real and imaginary spectrogram channels as independent inputs to a real-valued network \citep{jang21_interspeech,siuzdak2024vocos}. 
In contrast, cMRD uses complex-valued layers and operates directly on complex spectrogram inputs.
It comprises multiple sub-discriminators, each operating at a different STFT resolution.
During training, we apply the adversarial loss separately to the real and imaginary parts.
We also include a multi-period discriminator (MPD), shown in Figure~\ref{fig:model}(d), which consists of multiple sub-discriminators operating over different periods and processing reshaped waveform segments~\citep{NEURIPS2020_c5d73680}. 
Because the MPD operates at the waveform level, it remains a real-valued network. 
The overall training objective combines the adversarial losses from cMRD and MPD, along with feature matching and reconstruction losses. 
Full loss definitions and weights are provided in Appendix~\ref{app:objective}.

\subsection{Phase Quantization Layer}  
\label{sec:pq}
Complex-valued networks remain largely unexplored in terms of nonlinear transformations since any nonlinearity must jointly handle the real and imaginary components. 
We represent each Mel-spectrogram as a complex value by initializing the imaginary part to zero.
We then introduce a phase quantization layer that discretizes phase angles into a fixed set of levels. 
This provides a structured nonlinearity that preserves relative phase relationships and mitigates phase drift during training.
For a complex feature $z = r e^{i\theta}$, where $r \ge 0$ denotes the magnitude
and $\theta \in (-\pi, \pi]$ denotes the principal phase, the quantized phase
is defined as:

\begin{equation}
    \theta_q = \frac{2\pi}{N_q} \cdot \text{round} \left( \frac{N_q}{2\pi} \theta \right),
\end{equation}
where $N_q$ is the number of quantization levels.
The quantized complex value is reconstructed as
\begin{equation}
z_q = r e^{i\theta_q}.
\end{equation}

Quantizing the phase by mapping continuous angles to a fixed set of levels introduces inherent discontinuities that would normally block gradient propagation.
To preserve end-to-end differentiability, we adopt the straight-through
estimator (STE) \citep{bengio2013estimating}, in which the quantization
operation is applied in the forward pass, while its gradient is approximated
by an identity function during backpropagation.
This preserves gradient propagation through the phase quantization layer and improves optimization stability in practice.
Furthermore, by restricting phase values to a discrete set, phase quantization acts as a form of regularization: 
it limits unwarranted phase variability in intermediate representations and guides the network toward learning more coherent and structured phase patterns.

\subsection{Optimizing Complex Computation with Block Matrices}  
To improve efficiency in both the forward and backward passes, we reformulate CVNN operations as real-valued block-matrix multiplications.
In many autodifferentiation systems, complex-valued layers are implemented by explicitly tracking real and imaginary components as separate real-valued tensors.
This leads to redundant operations and inefficient memory access during both
the forward and backward passes.
We address this by adopting a block-wise formulation that represents complex values as structured pairs of real values and processes them jointly through unified matrix operations.
This approach reduces component-wise operations and enhances parallelism on modern GPU architectures by enabling matrix-based execution throughout the computational graph.
The forward complex operation can be expressed as:
\begin{align}
\begin{bmatrix}
\mathrm{Re}(z') \\
\mathrm{Im}(z')
\end{bmatrix}
&=
\begin{bmatrix}
W_r & -W_i \\
W_i & W_r
\end{bmatrix}
\begin{bmatrix}
x \\
y
\end{bmatrix},
\label{eq:forward}
\end{align}
where $z = x + i\,y$ (with $x$ and $y$ denoting the real and imaginary input vectors), $W = W_r + i\,W_i$ is the complex weight matrix (with $W_r$, $W_i$ its real and imaginary parts), and $z'$ is the resulting complex output.
The backward gradient computation uses the same block matrix structure:
\begin{align}
\begin{bmatrix}
\frac{\partial L}{\partial x} \\[0.5em]
\frac{\partial L}{\partial y}
\end{bmatrix}
&=
\begin{bmatrix}
W_r & -W_i \\
W_i & W_r
\end{bmatrix}^\top
\begin{bmatrix}
g_r \\
g_i
\end{bmatrix},
\label{eq:backward}
\end{align}
where $g_r$ and $g_i$ are the real and imaginary components of the gradient from the next layer.
This unified formulation is implemented for all parameterized CVNN layers via custom autograd functions.
It reduces the number of separate operations and improves parallelism on GPUs by replacing four independent real-valued multiplies with a single block-matrix multiply, thereby eliminating redundant computation and allowing more efficient gradient evaluation. 

\section{Results}
\subsection{Experimental Setup}
\label{experiments}
We train our model on the LibriTTS corpus \citep{zen19_interspeech}, using the \texttt{train-clean-100}, \texttt{train-clean-360}, and \texttt{train-other-500} subsets for training, and evaluating on \texttt{test-clean} and \texttt{test-other} sets.
All audio is sampled at 24 kHz. The STFT uses an FFT size of 1024, hop size of 256, and Hann window of length 1024.
Mel-spectrograms are computed with 100 Mel-bins and a maximum frequency of 12 kHz. 
We compare ComVo against several representative vocoders:
HiFi-GAN (v1) \citep{NEURIPS2020_c5d73680}, 
iSTFTNet \citep{9746713}, 
BigVGAN (base) \citep{lee2023bigvgan}, and 
Vocos \citep{siuzdak2024vocos}. 
For iSTFTNet, we use an open-source reimplementation, while the other models are trained using official code with recommended settings.
We evaluate using both subjective and objective metrics.
Subjective quality is assessed via mean opinion score (MOS), similarity MOS (SMOS), and comparison MOS (CMOS).
Objective metrics include UTMOS \citep{saeki22c_interspeech}, PESQ \citep{941023}, multi-resolution STFT (MR-STFT) error \citep{9053795}, periodicity RMSE, and V/UV F1 score \citep{morrison2022chunked}. 
Detailed explanations are provided in Appendix~\ref{app:baseline} and Appendix~\ref{app:metrics}.

\begin{table}[!t]
\caption{Objective and subjective evaluation on the LibriTTS dataset.}
\label{tab:main_results}
\centering
\resizebox{0.95\textwidth}{!}{%
\begin{tabular}{l|ccccc|cc}
\toprule
Model         & UTMOS $\uparrow$ & MR-STFT $\downarrow$ & PESQ $\uparrow$ & Periodicity $\downarrow$ & V/UV F1 $\uparrow$ & MOS $\uparrow$ & CMOS $\uparrow$ \\
\midrule
GT            & 3.8712 & -        & -   & -     & -     & 4.08 $\pm$ 0.04 & $0.14$ \\
\midrule
HiFi-GAN      & 3.3453 & 1.0455   & 2.9360 & 0.1554 & 0.9174 & 4.00 $\pm$ 0.05 & $-0.09$ \\
iSTFTNet      & 3.3591 & 1.1046   & 2.8136 & 0.1476 & 0.9243 & 3.98 $\pm$ 0.05 & $-0.04$ \\
BigVGAN       & 3.5197 & 0.8994   & 3.6122 & 0.1181 & 0.9418 & 4.05 $\pm$ 0.05 & $-0.05$ \\
Vocos         & 3.6025 & 0.8856   & 3.6266 & 0.1061 & 0.9522 & 4.05 $\pm$ 0.05 & $-0.02$ \\
ComVo         & \textbf{3.6901} & \textbf{0.8439} & \textbf{3.8239} & \textbf{0.0903} & \textbf{0.9609} & \textbf{4.07 $\pm$ 0.05} & $0$ \\
\bottomrule
\end{tabular} 
}
\end{table}

\begin{table}[!t]
\caption{Objective evaluation on the MUSDB18-HQ.}
\label{tab:ood_objective}
% \vskip 0.15in
\centering
\resizebox{0.62\textwidth}{!}{
\begin{tabular}{l|cccc}
\toprule
Model         & MR-STFT $\downarrow$ & PESQ $\uparrow$ & Periodicity $\downarrow$ & V/UV F1 $\uparrow$ \\
\midrule
HiFi-GAN      & 1.1909 & 2.3592 & 0.1804 & 0.9004 \\
iSTFTNet      & 1.2388 & 2.2357 & 0.1815 & 0.9102 \\
BigVGAN       & 0.9658 & 3.2391 & 0.1388 & 0.9340 \\
Vocos         & 0.9307 & 3.2785 & 0.1369 & 0.9361 \\
ComVo      & \textbf{0.8776} & \textbf{3.5220} & \textbf{0.1304} & \textbf{0.9384} \\
\bottomrule
\end{tabular} 
} 
\end{table}

\begin{table}[!t]
\caption{Subjective evaluation on the MUSDB18-HQ.}
\label{tab:ood_subjective}
\centering
\resizebox{0.90\textwidth}{!}{
\begin{tabular}{l|ccccc|c}
\toprule
Model         & Vocals & Drums & Bass & Others & Mixture & Average \\
\midrule
GT& 4.31 $\pm$ 0.11	& 4.25 $\pm$ 0.12 & 4.26 $\pm$ 0.12 & 4.29 $\pm$ 0.11 & 4.37 $\pm$ 0.11 & 4.29 $\pm$ 0.11 \\
\midrule
HiFi-GAN       & 3.83 $\pm$ 0.14	& 3.93 $\pm$ 0.13 & 3.43 $\pm$ 0.19 & 3.21 $\pm$ 0.19 & 3.60 $\pm$ 0.16 & 3.61 $\pm$ 0.16 \\
iSTFTNet      & 3.82 $\pm$ 0.14	& 4.03 $\pm$ 0.13 & 3.37 $\pm$ 0.18 & 3.17 $\pm$ 0.19 & 3.52 $\pm$ 0.17 & 3.59 $\pm$ 0.17 \\ 
BigVGAN       & \textbf{4.07 $\pm$ 0.12}	& \textbf{4.19 $\pm$ 0.12} & \underline{3.59} $\pm$ 0.17 & \underline{3.57} $\pm$ 0.15 & \underline{3.96} $\pm$ 0.12 & \underline{3.88} $\pm$ 0.14 \\ 
Vocos         & 4.04 $\pm$ 0.12	& 4.10 $\pm$ 0.13 & 3.58 $\pm$ 0.16 & 3.52 $\pm$ 0.17 & 3.87 $\pm$ 0.13 & 3.82 $\pm$ 0.14 \\
ComVo      & \underline{4.05} $\pm$ 0.12	& \underline{4.14} $\pm$ 0.12 & \textbf{3.60 $\pm$ 0.17} & \textbf{3.68 $\pm$ 0.16} & \textbf{3.98 $\pm$ 0.13} & \textbf{3.89 $\pm$ 0.14} \\
\bottomrule
\end{tabular} 
}
\end{table}

\subsection{Comparative Evaluation}
Table~\ref{tab:main_results} reports results on LibriTTS: ComVo achieves the highest objective scores among the baselines, and the corresponding MOS and CMOS are comparable to those of strong baseline systems.
Tables~\ref{tab:ood_objective} and~\ref{tab:ood_subjective} report results on MUSDB18-HQ~\citep{MUSDB18HQ}, an out-of-distribution audio dataset: 
ComVo achieves higher scores across all objective measures than the other models, and the corresponding subjective evaluations are comparable to strong baselines.
The SMOS evaluation shows that ComVo delivers competitive perceptual quality across individual source stems and mixture tracks, with its average scores typically at or near the top.
Taken together, these results indicate that an iSTFT-based model with complex-valued modeling consistently improves performance while maintaining the standard pipeline.

\begin{figure*}[!t]
\centering
\setlength{\tabcolsep}{3pt}
\renewcommand{\arraystretch}{1.0}
\begin{tabular}{>{\centering\arraybackslash}m{0.04\textwidth}|%
                >{\centering\arraybackslash}m{0.22\textwidth}%
                >{\centering\arraybackslash}m{0.22\textwidth}%
                >{\centering\arraybackslash}m{0.22\textwidth}%
                >{\centering\arraybackslash}m{0.22\textwidth}}
\toprule
& \textbf{$G_R D_R$} & \textbf{$G_C D_R$} & \textbf{$G_R D_C$} & \textbf{$G_C D_C$} \\
\midrule
\textbf{i} &
\includegraphics[width=\linewidth]{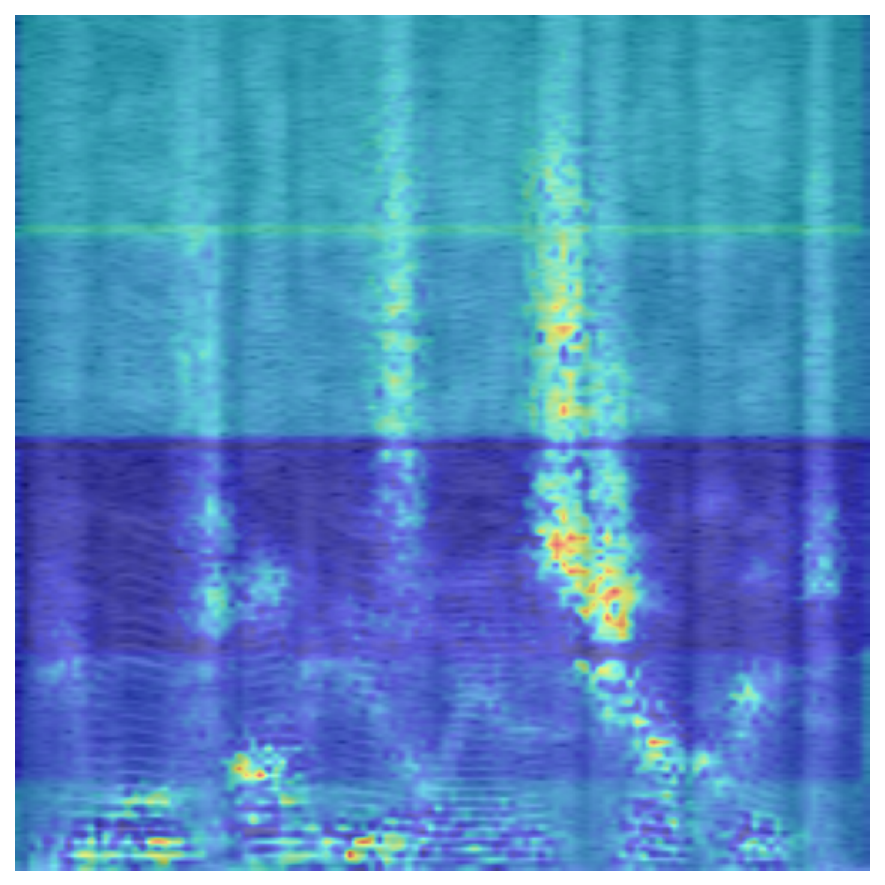} &
\includegraphics[width=\linewidth]{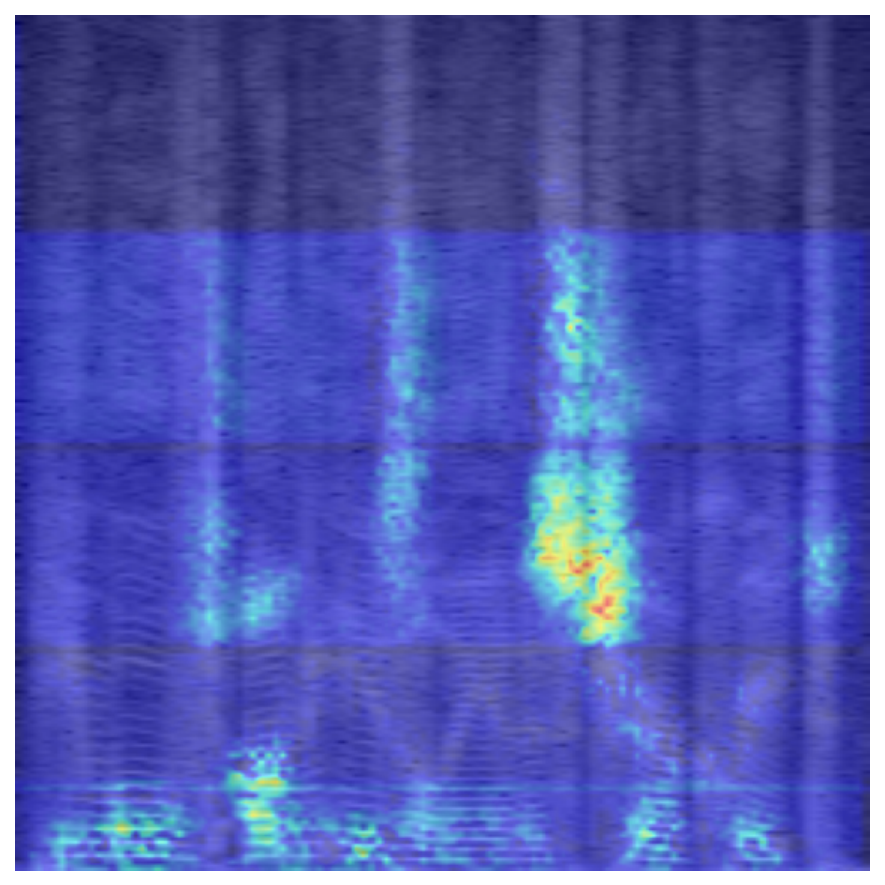} &
\includegraphics[width=\linewidth]{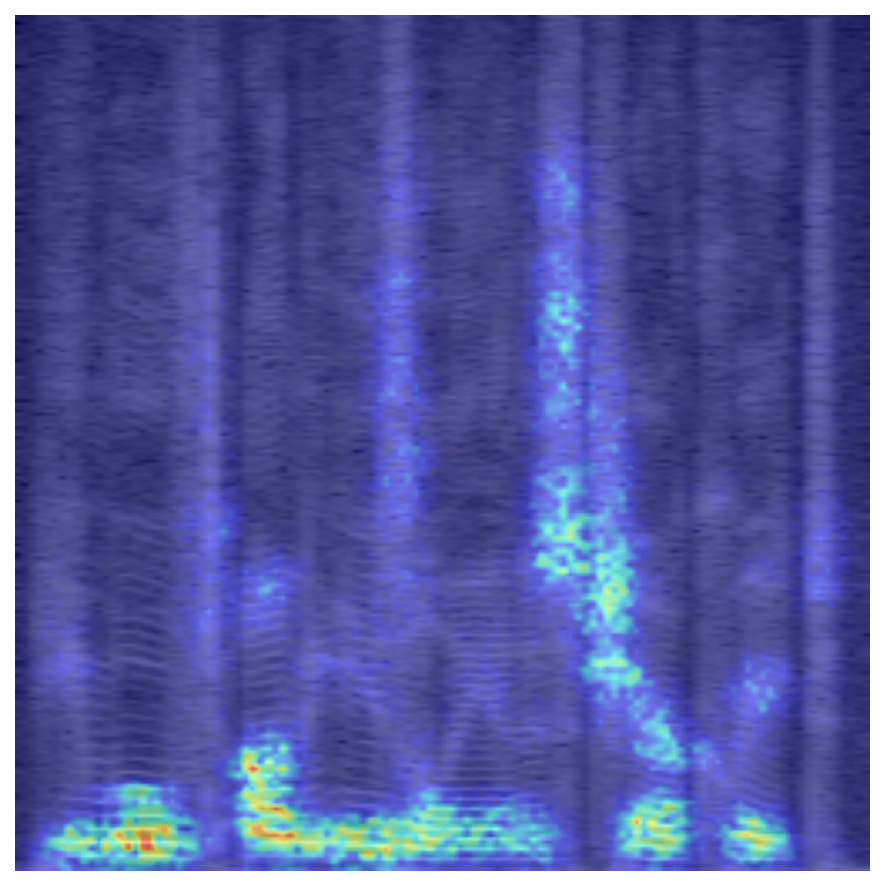} &
\includegraphics[width=\linewidth]{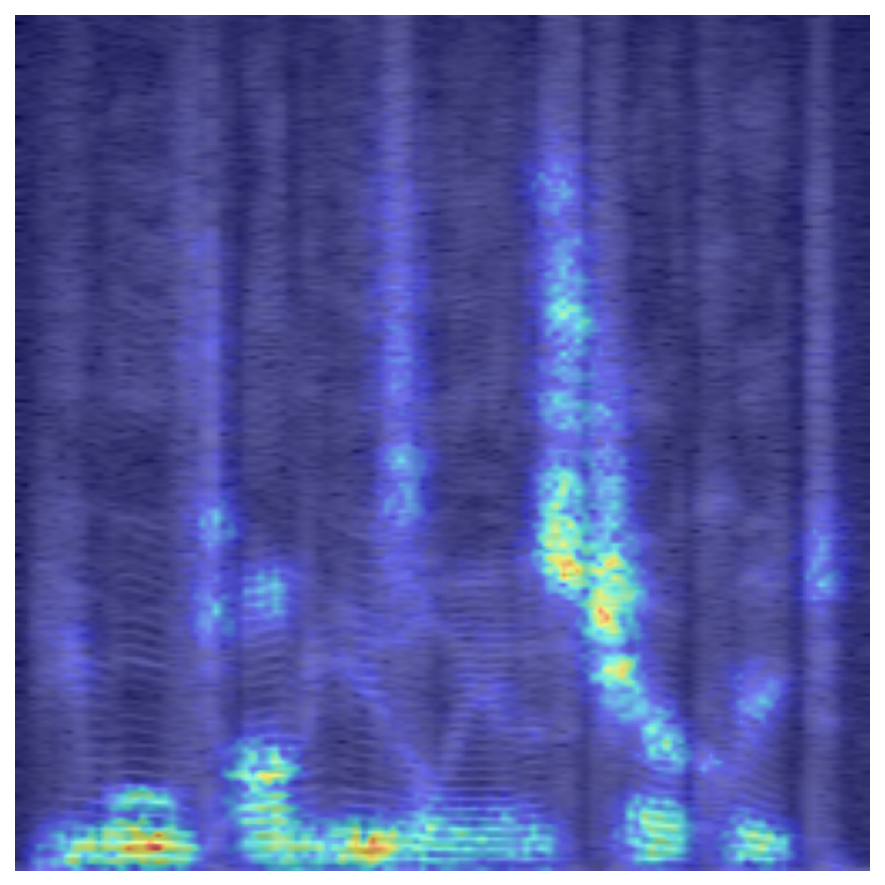} \\
\midrule
\textbf{ii} &
\includegraphics[width=\linewidth]{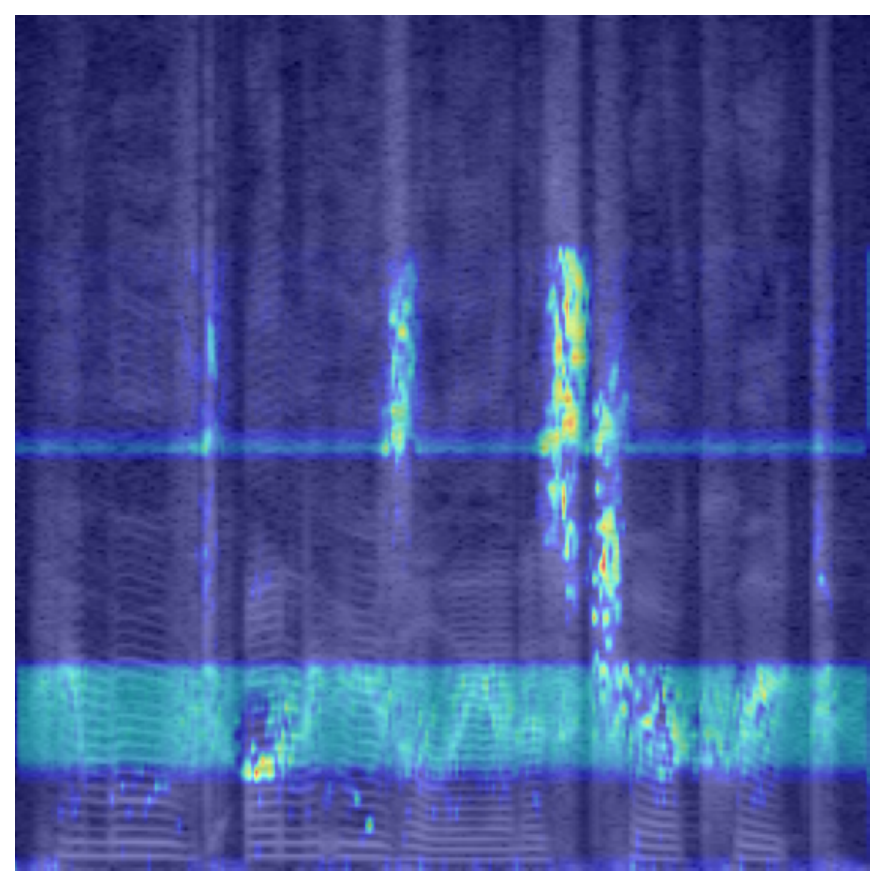} &
\includegraphics[width=\linewidth]{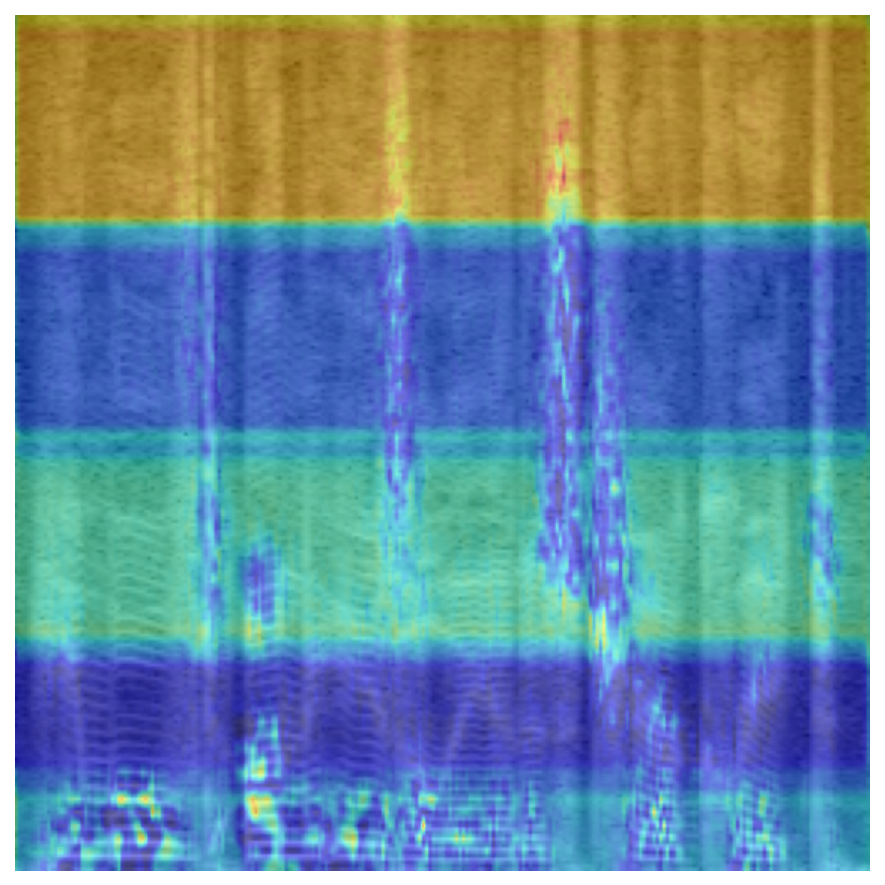} &
\includegraphics[width=\linewidth]{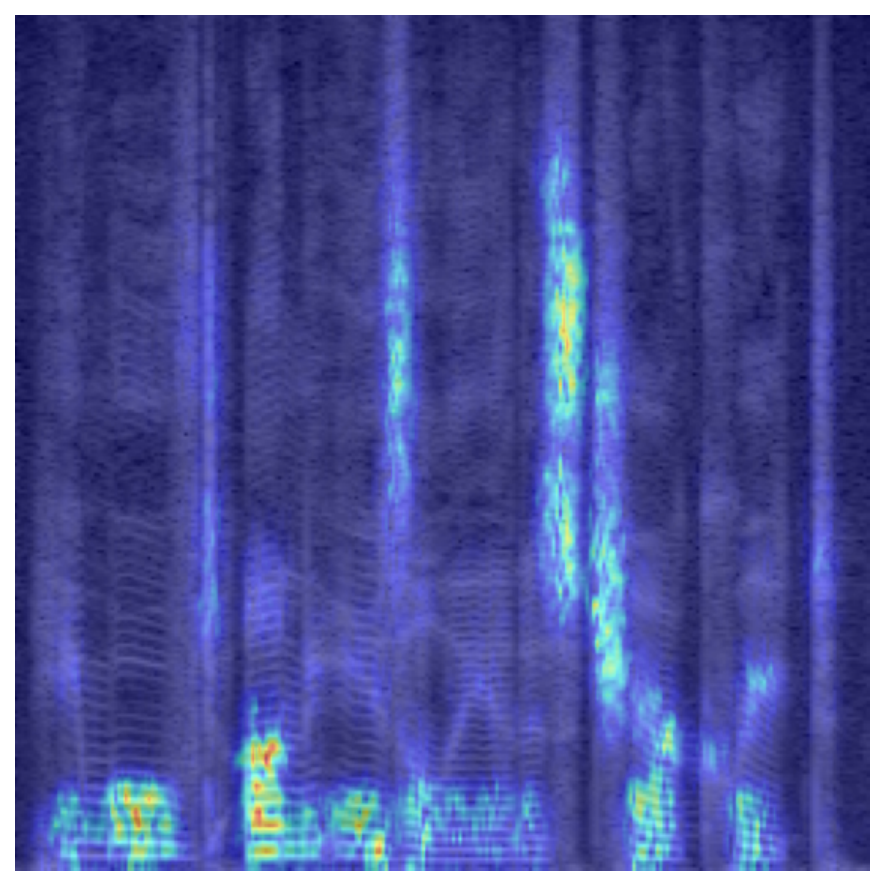} &
\includegraphics[width=\linewidth]{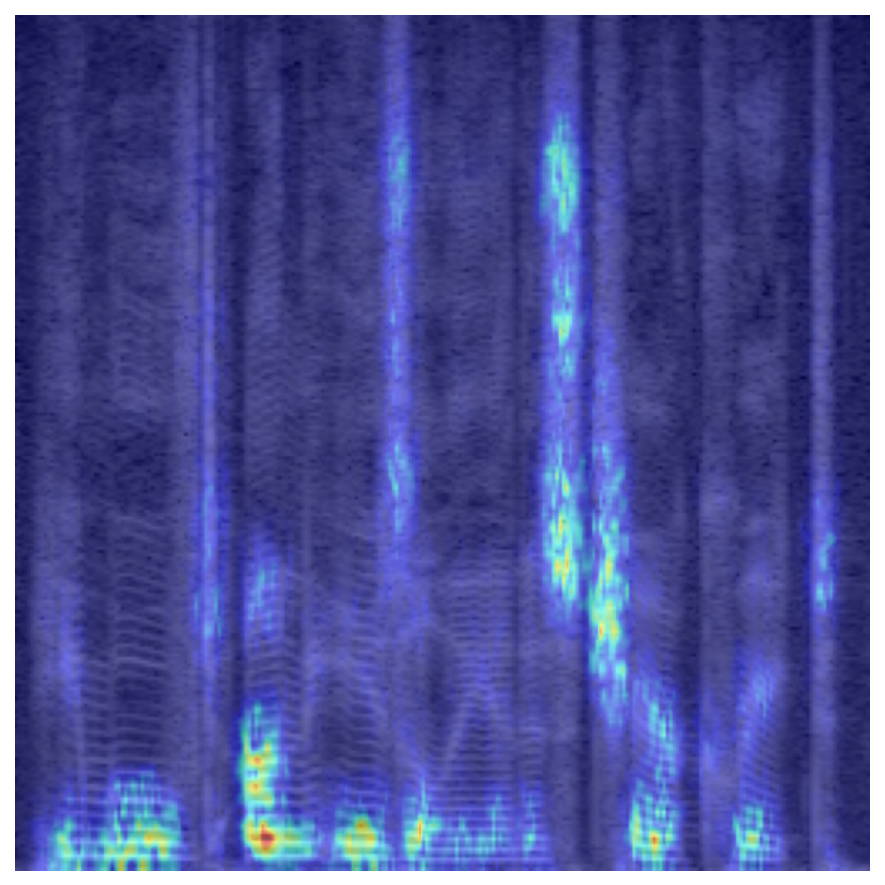} \\
\midrule
\textbf{iii} &
\includegraphics[width=\linewidth]{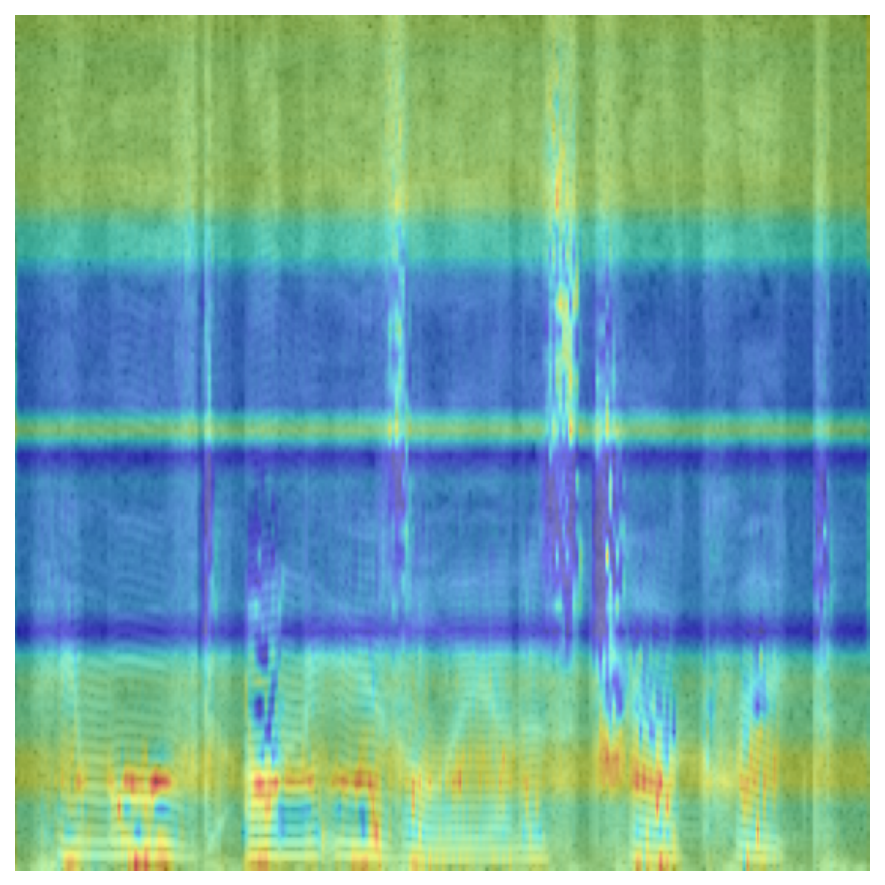} &
\includegraphics[width=\linewidth]{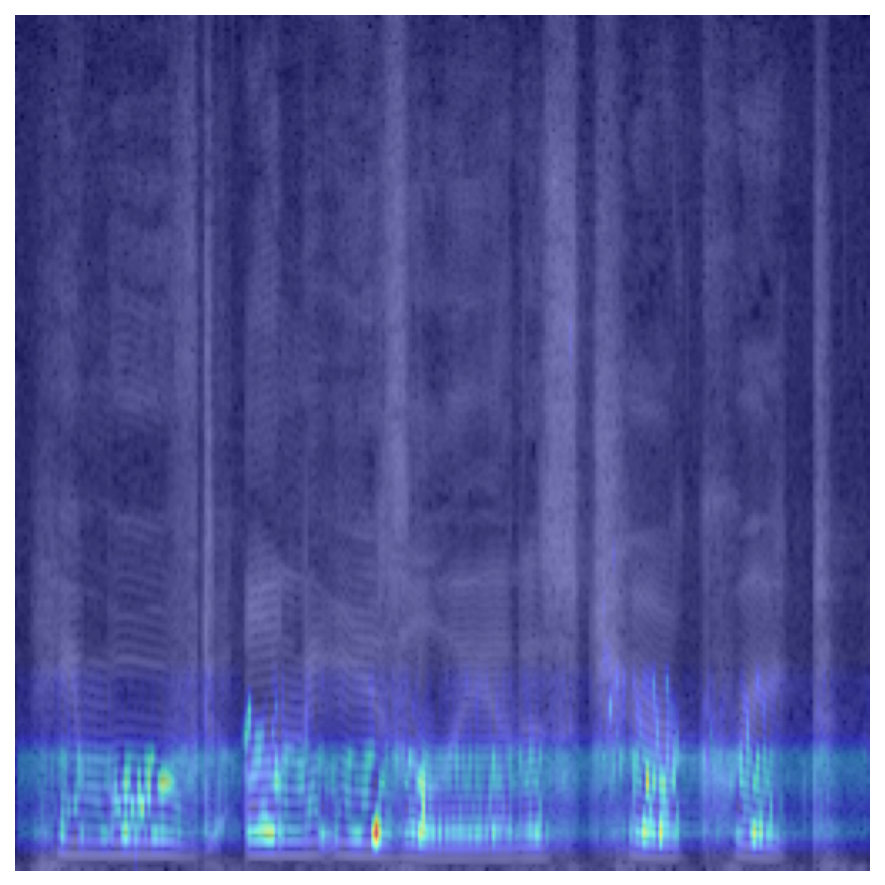} &
\includegraphics[width=\linewidth]{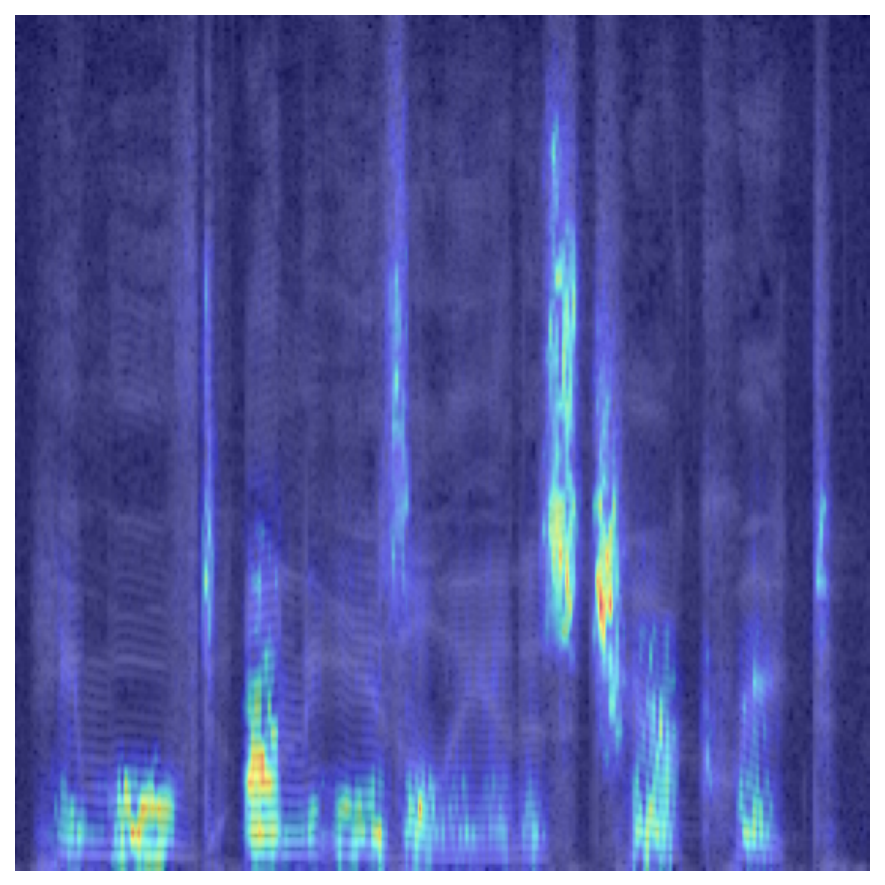} &
\includegraphics[width=\linewidth]{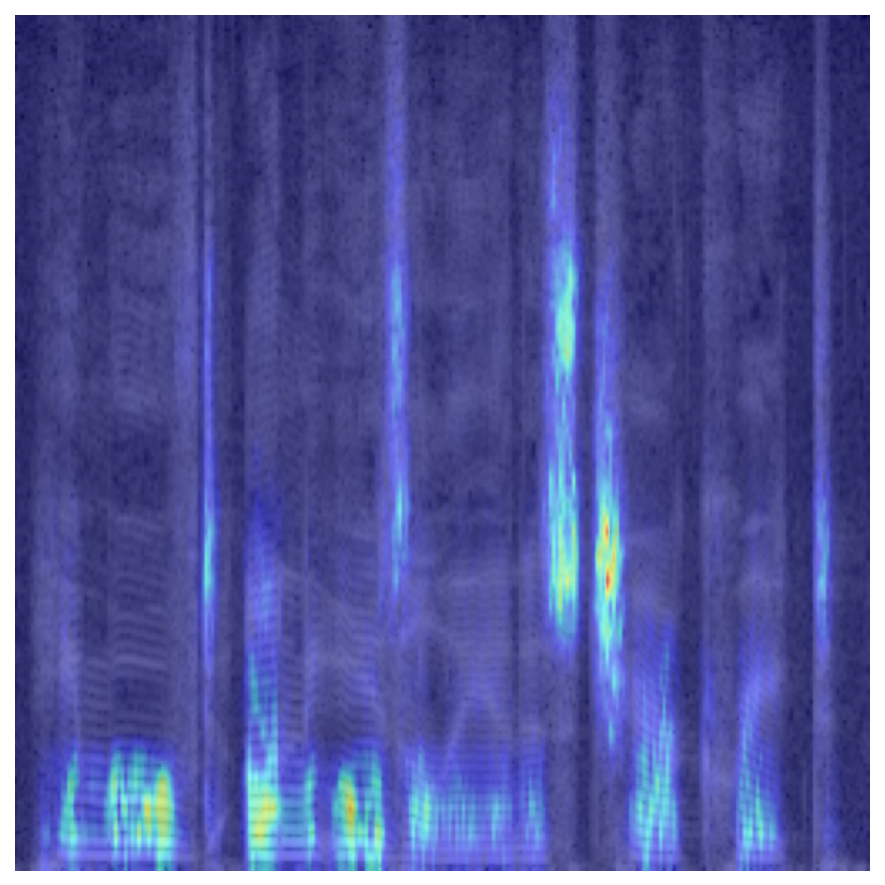} \\
\bottomrule
\end{tabular}
\caption{Grad-CAM comparison across generator-discriminator configurations. Each row corresponds to a cMRD sub-discriminator operating at a different STFT resolution (i, ii, iii).}
\label{fig:gradcam}
\vspace{-0.3cm}
\end{figure*}

\subsection{Impact of Complex-valued Modeling}
\label{sec:cvnn_ablation}

We assess the contribution of each discriminator component individually.
The MPD and MRD provide complementary forms of supervision: the MPD emphasizes periodic structure, while the MRD supplies multi-resolution spectral constraints.
To understand how each behaves on its own, we evaluate MPD-only, MRD-only, and cMRD-only configurations.
The MPD-only variant lacks spectral guidance and exhibits higher MR-STFT error.
The MRD-only variant attains low STFT-based errors but produces a lower UTMOS score, indicating that spectral constraints alone do not fully capture perceptual quality.
The cMRD-only model improves over the MRD-only baseline across all objective metrics, showing that the complex-valued discriminator provides a more effective constraint than its real-valued counterpart even when used alone.

We then extend the analysis to the full generator–discriminator combinations: $G_R D_R$, $G_C D_R$, $G_R D_C$, and $G_C D_C$, where $G_R$ and $G_C$ denote real-valued and complex-valued generators, and $D_R$ and $D_C$ denote real-valued and complex-valued discriminators.
To isolate the effect of complex-valued modeling, the phase-quantization layer is disabled for all configurations, and the MPD branch is kept active without modification.

Replacing only the generator ($G_R D_R \rightarrow G_C D_R$) consistently improves all objective metrics.
Replacing only the discriminator ($G_R D_R \rightarrow G_R D_C$) also yields measurable gains, particularly in MR-STFT error and PESQ.
The best performance is achieved when both the generator and discriminator operate in the complex domain ($G_C D_C$), confirming the effectiveness of complex-domain modeling for iSTFT-based waveform generation.

For qualitative analysis, we visualize Grad-CAM \citep{Selvaraju_2017_ICCV} activations of the discriminator in Figure~\ref{fig:gradcam}.  
Each row in the figure corresponds to a sub-discriminator index (i, ii, iii), and each column corresponds to one of the generator-discriminator configurations.
In the configurations with a real-valued MRD ($G_R D_R$ and $G_C D_R$), the attention maps are diffuse and poorly aligned with speech-relevant spectral structures.
In contrast, in the configurations with a cMRD ($G_R D_C$ and $G_C D_C$), the highlighted regions consistently trace structured spectral patterns across all sub-discriminators.
These results indicate that complex-valued discriminators provide more precise spectral feedback to the generator, helping it better match perceptually important features and ultimately improving synthesis quality, as also reflected in the ablation metrics.

\begin{table}[!t]
\caption{Ablation study comparing real-valued and complex-valued architectures.}
\label{tab:cvnn_ablation}
% \vskip 0.15in
\centering
\resizebox{0.73\textwidth}{!}{%
\begin{tabular}{l|cccccc}
\toprule
Model & UTMOS $\uparrow$ & MR-STFT $\downarrow$ & PESQ $\uparrow$ & Periodicity $\downarrow$ & V/UV F1 $\uparrow$\\
\midrule
 MPD only  &  3.6357 &  0.8522 &  3.7670 &  0.0942 &  0.9613 \\
 MRD only  &  2.8338 &  0.8442 &  3.9868 &  0.0870 &  0.9610 \\
 cMRD only &  2.9285 &  0.8398 &  4.0149 &  0.0859 &  0.9635 \\
\midrule

$G_R D_R$ & 3.6025 & 0.8856 & 3.6266 & 0.1061 & 0.9522 \\
$G_R D_C$ & 3.5930 & 0.8679 & 3.6399 & 0.1060 & 0.9497 \\
$G_C D_R$ & 3.6452 & 0.8597 & 3.7375 & 0.0978 & 0.9567 \\
$G_C D_C$ & 3.6646 & 0.8435 & 3.7756 & 0.0915 & 0.9625 \\
\bottomrule
\end{tabular}
}
\vspace{-0.3cm}
\end{table}

\subsection{Effect of Phase Quantization}
\label{sec:phase_quantization}
Table~\ref{tab:phase_ablation} shows that adding a phase quantization layer yields clear benefits in perceptual quality, despite only a minor trade-off in reconstruction fidelity. 
The model without phase quantization ($N_q=0$) achieves the lowest MR-STFT error, but a moderate quantization level ($N_q=128$) smooths out phase fluctuations, resulting in higher UTMOS and PESQ scores and fewer periodicity artifacts, with only a small increase in MR-STFT error. 
Using finer quantization (e.g., $N_q=256$, $N_q=512$) can further boost perceptual metrics, but with diminishing returns and a slight degradation in reconstruction accuracy.
Overall, phase quantization acts as an effective regularizer:
it enhances listening quality while only modestly affecting spectral fidelity, with $N_q=128$ providing the best trade-off in our setup.

\begin{table}[!t]
\caption{Ablation on phase quantization levels. $N_q$ denotes the number of quantization levels.}
\label{tab:phase_ablation}
\centering
\resizebox{0.8\textwidth}{!}{
\begin{tabular}{c|cccccc}
\toprule
$N_q$ Quantization & UTMOS $\uparrow$ & MR-STFT $\downarrow$ & PESQ $\uparrow$ & Periodicity $\downarrow$ & V/UV F1 $\uparrow$\\
\midrule
0   & \underline{3.6646} & \textbf{0.8435} & 3.7756 & 0.0915 & \textbf{0.9625} \\
\midrule
128 & \textbf{3.6901} & \underline{0.8439} & \underline{3.8239} & \underline{0.0903} & 0.9609 \\
256 & 3.6423 & 0.8466 & 3.8127 & 0.0926 & 0.9597 \\
512 & 3.6412 & 0.8489 & \textbf{3.8248} & \textbf{0.0896} & \underline{0.9613} \\
\bottomrule
\end{tabular}
}
\end{table}

\begin{table}[!t]
\caption{Comparison of standard PyTorch and refined implementations.
} 
\centering
\label{tab:refine_results}
\resizebox{0.8\textwidth}{!}{%
\begin{tabular}{l|cccccc}
\toprule
Implementation & MR-STFT $\downarrow$ & GPU xRT $\uparrow$ & Training Time & Nodes (Gen / cMRD)\\
\midrule
Native PyTorch  & 0.8465 & \textbf{702.26} & 183 hrs & 5686 / 4248 \\
Block-matrix    & \textbf{0.8435} & 696.91 & \textbf{138 hrs} & \textbf{2547 / 1404} \\
\bottomrule
\end{tabular} 
} 
\vspace{-0.3cm}
\end{table}

\subsection{Block-matrix Computation Scheme}
\label{exp_block}
In this section, we evaluate the efficiency and graph-complexity benefits of our block-matrix computation scheme.  
Table~\ref{tab:refine_results} reports the comparative results. 
It shows that our block-matrix implementation achieves performance comparable to PyTorch’s native complex operations in terms of MR-STFT reconstruction error.
While PyTorch’s optimized complex kernels yield slightly faster forward-pass throughput, our overall training time is substantially shorter. 
Specifically, we reduce the number of backward graph nodes in the generator by over 55\% and in the discriminator’s cMRD by nearly 67\%, resulting in a 25\% reduction in training time. 
This improvement arises primarily from the backward pass: examining the gradient computation graphs reveals that our method dramatically lowers the node count compared to PyTorch’s default approach of separately tracking real and imaginary components.
By replacing four independent real-valued multiplications with a simple channel concatenation and a single matrix multiplication, we eliminate redundant operations and significantly accelerate gradient computation, all without sacrificing model fidelity.

\subsection{Evaluation in Text-to-speech Pipeline}
\label{sec:tts}
We further evaluate each model in a text-to-speech (TTS) pipeline by pairing it with an acoustic model. 
In particular, we use Matcha-TTS~\citep{10448291} as the acoustic model to generate Mel-spectrograms from text, then pass those spectrograms to each model. 
Matcha-TTS is trained on LibriTTS, and each model is trained independently on LibriTTS and connected to the Matcha-TTS outputs without additional fine-tuning.
Table~\ref{tab:tts_results} reports the MOS, UTMOS, and CMOS for the TTS pipeline evaluation. 
ComVo achieves a MOS that matches the top score among the compared models, and it attains the highest UTMOS.
This indicates that ComVo reliably converts the predicted spectrograms into high-quality waveforms within the TTS setting.

\subsection{Computational Analysis}
Table~\ref{tab:computational_cost} compares the inference throughput and memory usage of each model under a common setup (batch size~1, no hardware-specific optimizations).
HiFi-GAN and BigVGAN are upsampling-based models, whereas iSTFTNet, Vocos, and ComVo synthesize via frame-level iSTFT. 
The upsampling-based models exhibit the lowest throughput (lower xRT, indicating slower generation), while the iSTFT-based models run significantly faster.
Among them, Vocos achieves the highest throughput.
ComVo’s throughput (xRT) lies within the range of the other iSTFT-based models. 
However, its memory footprint is higher than the real-valued iSTFT baselines: with a complex type, each weight is stored as a real–imaginary pair, so at the same precision the per-parameter memory is roughly doubled for a fixed parameter count.

To test whether the improvements stem merely from the larger memory footprint of complex types, we trained a real-valued model with twice the parameter count to match the complex model’s memory and compared cost--quality trade-offs. 
The results are reported in Table~\ref{tab:cvnn_rvnn}.
We compare three settings: the baseline real-valued model ($G_R D_R$), a widened real-valued model with roughly $2\times$ parameters (denoted $G_R D_R$ 2$\times$), and a complex-valued model ($G_C D_R$). 
The discriminator is identical across all settings.
$G_C D_R$ and $G_R D_R$ 2$\times$ have comparable memory footprints.
As expected, $G_R D_R$ 2$\times$ improves objective metrics relative to $G_R D_R$. 
In fact, $G_C D_R$ exceeds the widened model across all metrics despite a similar memory cost.
Taken together, Tables~\ref{tab:computational_cost} and~\ref{tab:cvnn_rvnn} indicate that modeling real–imaginary correlations with CVNNs provides larger quality gains than simply scaling real-valued models.

\begin{table}[!t]
  \centering
  \begin{minipage}{0.45\textwidth}
        \centering
        \caption{UTMOS, MOS, and CMOS comparison in the TTS pipeline.} 
        \label{tab:tts_results}
        \resizebox{1.0\linewidth}{!}{%
        \begin{tabular}{l|ccc}
        \toprule
        Model & UTMOS $\uparrow$ & MOS $\uparrow$ & CMOS $\uparrow$ \\
        \midrule
        HiFi-GAN & 3.2233 & 3.85 $\pm$ 0.05 & $-0.22$ \\
        iSTFTNet & 3.2951 & 3.89 $\pm$ 0.05 & $-0.15$ \\
        BigVGAN  & 3.3022 & 3.92 $\pm$ 0.05 & $-0.06$ \\
        Vocos    & 3.4357 & 3.91 $\pm$ 0.05 & $-0.06$ \\
        ComVo    & \textbf{3.4403} & \textbf{3.92 $\pm$ 0.05} & $0$ \\
        \bottomrule
        \end{tabular}
        }
  \end{minipage}
  \hfill
  \begin{minipage}{0.50\textwidth}
        \centering
        \caption{Comparison of computational cost and inference latency.} 
        \label{tab:computational_cost}
        \resizebox{1.0\linewidth}{!}{%
        \begin{tabular}{l|ccc}
        \toprule
        Model & Param (M) & Memory (MB) & GPU xRT $\uparrow$ \\
        \midrule
        HiFi-GAN      & 14.00 & 53.40 & 259.08 \\
        iSTFTNet      & 13.33 & 50.83 & 402.21 \\
        BigVGAN       & 14.02 & 53.46 & 158.07 \\
        Vocos         & 13.54 & 51.62 & 4657.65 \\
        ComVo         & 13.28 & 101.24 & 819.02 \\
        \bottomrule
        \end{tabular}
        }
  \end{minipage}
\end{table}

\begin{table}[!t]
\caption{Objective evaluation and cost comparison: complex modeling vs. parameter scaling.
}
\label{tab:cvnn_rvnn}
% \vskip 0.15in
\centering
\resizebox{1.0\textwidth}{!}{%
\begin{tabular}{l|cc|cccccc}
\toprule
Model & Params. (M) & Memory (MB) & UTMOS $\uparrow$ & MR-STFT $\downarrow$ & PESQ $\uparrow$ & Periodicity $\downarrow$ & V/UV F1 $\uparrow$ \\
\midrule
$G_R D_R$ & 13.54 & 51.62 & 3.6025 & 0.8856 & 3.6266 & 0.1061 & 0.9522 \\
$G_R D_R$ 2 $\times$ & 27.05 & 103.19 & 3.6164 & 0.8622 & 3.6336 & 0.1055 & 0.9524 \\
$G_C D_R$ & 13.28 & 101.24 & \textbf{3.6452} & \textbf{0.8597} & \textbf{3.7375} & \textbf{0.0978} & \textbf{0.9567} \\
\bottomrule
\end{tabular}
}
\end{table}

\section{Limitations}
ComVo integrates complex-valued networks into an iSTFT-based vocoder. 
To keep the implementation straightforward, we adopt split-style designs. 
Concretely, we apply component-wise hinge losses to the real and imaginary outputs of cMRD, and we use split GELU within the ConvNeXt backbone.
We will explore more advanced designs for these components in future work.
The block-matrix formulation accelerates training, but computational overhead remains high because complex layers store and process paired real and imaginary values. 
Empirically, multi-GPU Distributed Data Parallel experiments showed under-optimized performance for complex parameters in our current training setup and occasional numerical issues; accordingly, we report single-GPU results.
With better multi-GPU optimization and broader design exploration, larger-scale studies should be feasible and can further catalyze research on CVNNs for speech generation.

\section{Conclusion}
\label{conclusion}
We presented ComVo, a vocoder that integrates CVNNs into both the generator and the discriminator, establishing a complex-domain adversarial framework for iSTFT-based waveform generation. 
By modeling the real and imaginary components jointly, our method addresses the structural mismatches in conventional real-valued processing of complex spectrograms. 
We also introduced a phase quantization layer as an inductive bias and a block-matrix formulation that simplifies computation graphs and accelerates training.
ComVo delivered higher synthesis quality than comparable real-valued baselines. 
In addition, the block-matrix formulation reduced training time by approximately 25\%.
Future work will extend this framework beyond adversarial training to other generative paradigms (e.g., diffusion or flow-matching) and explore richer complex-domain activations and losses.

\section*{Acknowledgments}
This work was partly supported by Institute of Information \& communications Technology Planning \& Evaluation(IITP) grant funded by the Korea government(MSIT) (No. RS-2019-II190079, Artificial Intelligence Graduate School Program (Korea University), IITP-2026-RS-2025-02304828, Artificial Intelligence Star Fellowship Support Program to nurture the best talents and No. RS-2024-00457882, AI Research Hub Project). 

\bibliography{iclr2026_conference}

@inproceedings{NEURIPS2020_c5d73680,
 author = {Kong, Jungil and Kim, Jaehyeon and Bae, Jaekyoung},
 booktitle = {Advances in Neural Information Processing Systems},
 title = {HiFi-GAN: Generative Adversarial Networks for Efficient and High Fidelity Speech Synthesis},
 year = {2020}
}

@inproceedings{jang21_interspeech,
  title     = {UnivNet: A Neural Vocoder with Multi-Resolution Spectrogram Discriminators for High-Fidelity Waveform Generation},
  author    = {Won Jang and Dan Lim and Jaesam Yoon and Bongwan Kim and Juntae Kim},
  year      = {2021},
  booktitle = {Interspeech 2021},
  pages     = {2207--2211},
  doi       = {10.21437/Interspeech.2021-1016},
  issn      = {2958-1796},
}

@inproceedings{
lee2023bigvgan,
title={{BigVGAN: A Universal Neural Vocoder with Large-Scale Training}},
author={Sang-Gil Lee and Wei Ping and Boris Ginsburg and Bryan Catanzaro and Sungroh Yoon},
booktitle={The Eleventh International Conference on Learning Representations},
year={2023},
}

@inproceedings{
siuzdak2024vocos,
title={Vocos: Closing the gap between time-domain and Fourier-based neural vocoders for high-quality audio synthesis},
author={Hubert Siuzdak},
booktitle={The Twelfth International Conference on Learning Representations},
year={2024},
}

@InProceedings{Liu_2022_CVPR,
    author    = {Liu, Zhuang and Mao, Hanzi and Wu, Chao-Yuan and Feichtenhofer, Christoph and Darrell, Trevor and Xie, Saining},
    title     = {A ConvNet for the 2020s},
    booktitle = {Proceedings of the IEEE/CVF Conference on Computer Vision and Pattern Recognition (CVPR)},
    month     = {June},
    year      = {2022},
    pages     = {11976-11986}
}

@InProceedings{Vasudeva_2022_WACV,
    author    = {Vasudeva, Bhavya and Deora, Puneesh and Bhattacharya, Saumik and Pradhan, Pyari Mohan},
    title     = {Compressed Sensing MRI Reconstruction With Co-VeGAN: Complex-Valued Generative Adversarial Network},
    booktitle = {Proceedings of the IEEE/CVF Winter Conference on Applications of Computer Vision (WACV)},
    month     = {January},
    year      = {2022},
    pages     = {672-681}
}

@inproceedings{zen19_interspeech,
  title     = {LibriTTS: A Corpus Derived from LibriSpeech for Text-to-Speech},
  author    = {Heiga Zen and Viet Dang and Rob Clark and Yu Zhang and Ron J. Weiss and Ye Jia and Zhifeng Chen and Yonghui Wu},
  year      = {2019},
  booktitle = {Interspeech 2019},
  pages     = {1526--1530},
  doi       = {10.21437/Interspeech.2019-2441},
  issn      = {2958-1796},
}

@INPROCEEDINGS{941023,
  author={Rix, A.W. and Beerends, J.G. and Hollier, M.P. and Hekstra, A.P.},
  booktitle={2001 IEEE International Conference on Acoustics, Speech, and Signal Processing. Proceedings (Cat. No.01CH37221)}, 
  title={Perceptual evaluation of speech quality (PESQ)-a new method for speech quality assessment of telephone networks and codecs}, 
  year={2001},
  volume={2},
  number={},
  pages={749-752 vol.2},
  doi={10.1109/ICASSP.2001.941023}}

@inproceedings{saeki22c_interspeech,
  title     = {UTMOS: UTokyo-SaruLab System for VoiceMOS Challenge 2022},
  author    = {Takaaki Saeki and Detai Xin and Wataru Nakata and Tomoki Koriyama and Shinnosuke Takamichi and Hiroshi Saruwatari},
  year      = {2022},
  booktitle = {Interspeech 2022},
  pages     = {4521--4525},
  doi       = {10.21437/Interspeech.2022-439},
  issn      = {2958-1796},
}

@inproceedings{
morrison2022chunked,
title={Chunked Autoregressive {GAN} for Conditional Waveform Synthesis},
author={Max Morrison and Rithesh Kumar and Kundan Kumar and Prem Seetharaman and Aaron Courville and Yoshua Bengio},
booktitle={The Tenth International Conference on Learning Representations},
year={2022},
}

@inproceedings{kaneko23_interspeech,
  title     = {iSTFTNet2: Faster and More Lightweight iSTFT-Based Neural Vocoder Using 1D-2D CNN},
  author    = {Takuhiro Kaneko and Hirokazu Kameoka and Kou Tanaka and Shogo Seki},
  year      = {2023},
  booktitle = {Interspeech 2023},
  pages     = {4369--4373},
  doi       = {10.21437/Interspeech.2023-1726},
  issn      = {2958-1796},
}

@inproceedings{NEURIPS2019_6804c9bc,
 author = {Kumar, Kundan and Kumar, Rithesh and de Boissiere, Thibault and Gestin, Lucas and Teoh, Wei Zhen and Sotelo, Jose and de Br\'{e}bisson, Alexandre and Bengio, Yoshua and Courville, Aaron C},
 booktitle = {Advances in Neural Information Processing Systems},
 editor = {H. Wallach and H. Larochelle and A. Beygelzimer and F. d\textquotesingle Alch\'{e}-Buc and E. Fox and R. Garnett},
 pages = {},
 publisher = {Curran Associates, Inc.},
 title = {MelGAN: Generative Adversarial Networks for Conditional Waveform Synthesis},
 volume = {32},
 year = {2019}
}

@inproceedings{
kong2021diffwave,
title={DiffWave: A Versatile Diffusion Model for Audio Synthesis},
author={Zhifeng Kong and Wei Ping and Jiaji Huang and Kexin Zhao and Bryan Catanzaro},
booktitle={The Ninth International Conference on Learning Representations},
year={2021},
}

@INPROCEEDINGS{9053795,
  author={Yamamoto, Ryuichi and Song, Eunwoo and Kim, Jae-Min},
  booktitle={ICASSP 2020 - 2020 IEEE International Conference on Acoustics, Speech and Signal Processing (ICASSP)}, 
  title={Parallel Wavegan: A Fast Waveform Generation Model Based on Generative Adversarial Networks with Multi-Resolution Spectrogram}, 
  year={2020},
  volume={},
  number={},
  pages={6199-6203},
  doi={10.1109/ICASSP40776.2020.9053795}}

@INPROCEEDINGS{9746713,
  author={Kaneko, Takuhiro and Tanaka, Kou and Kameoka, Hirokazu and Seki, Shogo},
  booktitle={ICASSP 2022 - 2022 IEEE International Conference on Acoustics, Speech and Signal Processing (ICASSP)}, 
  title={ISTFTNET: Fast and Lightweight Mel-Spectrogram Vocoder Incorporating Inverse Short-Time Fourier Transform}, 
  year={2022},
  volume={},
  number={},
  pages={6207-6211},
  doi={10.1109/ICASSP43922.2022.9746713}}

@inproceedings{
trabelsi2018deep,
title={Deep Complex Networks},
author={Chiheb Trabelsi and Olexa Bilaniuk and Ying Zhang and Dmitriy Serdyuk and Sandeep Subramanian and Joao Felipe Santos and Soroush Mehri and Negar Rostamzadeh and Yoshua Bengio and Christopher J Pal},
booktitle={The Sixth International Conference on Learning Representations},
year={2018},
}

@INPROCEEDINGS{8553396,
  author={Oyamada, Keisuke and Kameoka, Hirokazu and Kaneko, Takuhiro and Tanaka, Kou and Hojo, Nobukatsu and Ando, Hiroyasu},
  booktitle={2018 26th European Signal Processing Conference (EUSIPCO)}, 
  title={Generative adversarial network-based approach to signal reconstruction from magnitude spectrogram}, 
  year={2018},
  volume={},
  number={},
  pages={2514-2518},
  doi={10.23919/EUSIPCO.2018.8553396}}

@inproceedings{neekhara19_interspeech,
  title     = {Expediting TTS Synthesis with Adversarial Vocoding},
  author    = {Paarth Neekhara and Chris Donahue and Miller Puckette and Shlomo Dubnov and Julian McAuley},
  year      = {2019},
  booktitle = {Interspeech 2019},
  pages     = {186--190},
  doi       = {10.21437/Interspeech.2019-3099},
  issn      = {2958-1796},
}

@inproceedings{NEURIPS2020_9873eaad,
 author = {Gritsenko, Alexey and Salimans, Tim and van den Berg, Rianne and Snoek, Jasper and Kalchbrenner, Nal},
 booktitle = {Advances in Neural Information Processing Systems},
 editor = {H. Larochelle and M. Ranzato and R. Hadsell and M.F. Balcan and H. Lin},
 pages = {13062--13072},
 publisher = {Curran Associates, Inc.},
 title = {A Spectral Energy Distance for Parallel Speech Synthesis},
 volume = {33},
 year = {2020}
}

@InProceedings{pmlr-v80-oord18a,
  title = 	 {Parallel {W}ave{N}et: Fast High-Fidelity Speech Synthesis},
  author =       {van den Oord, Aaron and Li, Yazhe and Babuschkin, Igor and Simonyan, Karen and Vinyals, Oriol and Kavukcuoglu, Koray and van den Driessche, George and Lockhart, Edward and Cobo, Luis and Stimberg, Florian and Casagrande, Norman and Grewe, Dominik and Noury, Seb and Dieleman, Sander and Elsen, Erich and Kalchbrenner, Nal and Zen, Heiga and Graves, Alex and King, Helen and Walters, Tom and Belov, Dan and Hassabis, Demis},
  booktitle = 	 {Proceedings of the 35th International Conference on Machine Learning},
  pages = 	 {3918--3926},
  year = 	 {2018},
  editor = 	 {Dy, Jennifer and Krause, Andreas},
  volume = 	 {80},
  series = 	 {Proceedings of Machine Learning Research},
  month = 	 {10--15 Jul},
  publisher =    {PMLR}
}

@InProceedings{pmlr-v119-ping20a,
  title = 	 {{W}ave{F}low: A Compact Flow-based Model for Raw Audio},
  author =       {Ping, Wei and Peng, Kainan and Zhao, Kexin and Song, Zhao},
  booktitle = 	 {Proceedings of the 37th International Conference on Machine Learning},
  pages = 	 {7706--7716},
  year = 	 {2020},
  editor = 	 {III, Hal Daumé and Singh, Aarti},
  volume = 	 {119},
  series = 	 {Proceedings of Machine Learning Research},
  month = 	 {13--18 Jul},
  publisher =    {PMLR}
}

@inproceedings{NEURIPS2020_a1c3ae6c,
 author = {Lee, Sang-Gil and Kim, Sungwon and Yoon, Sungroh},
 booktitle = {Advances in Neural Information Processing Systems},
 editor = {H. Larochelle and M. Ranzato and R. Hadsell and M.F. Balcan and H. Lin},
 pages = {14058--14067},
 publisher = {Curran Associates, Inc.},
 title = {NanoFlow: Scalable Normalizing Flows with Sublinear Parameter Complexity},
 volume = {33},
 year = {2020}
}

@inproceedings{
lee2022priorgrad,
title={PriorGrad: Improving Conditional Denoising Diffusion Models with Data-Dependent Adaptive Prior},
author={Sang-Gil Lee and Heeseung Kim and Chaehun Shin and Xu Tan and Chang Liu and Qi Meng and Tao Qin and Wei Chen and Sungroh Yoon and Tie-Yan Liu},
booktitle={The Tenth International Conference on Learning Representations},
year={2022},
}

@inproceedings{
chen2021wavegrad,
title={WaveGrad: Estimating Gradients for Waveform Generation},
author={Nanxin Chen and Yu Zhang and Heiga Zen and Ron J Weiss and Mohammad Norouzi and William Chan},
booktitle={The Ninth International Conference on Learning Representations},
year={2021},
}

@ARTICLE{1164317,
  author={Griffin, D. and Jae Lim},
  journal={IEEE Transactions on Acoustics, Speech, and Signal Processing}, 
  title={Signal estimation from modified short-time Fourier transform}, 
  year={1984},
  volume={32},
  number={2},
  pages={236-243},
  doi={10.1109/TASSP.1984.1164317}}

@INPROCEEDINGS{10446058,
  author={Liu, Haocheng and Baoueb, Teysir and Fontaine, Mathieu and Le Roux, Jonathan and Richard, Gaël},
  booktitle={ICASSP 2024 - 2024 IEEE International Conference on Acoustics, Speech and Signal Processing (ICASSP)}, 
  title={GLA-GRAD: A Griffin-Lim Extended Waveform Generation Diffusion Model}, 
  year={2024},
  volume={},
  number={},
  pages={11611-11615},
  doi={10.1109/ICASSP48485.2024.10446058}}

@article{hendrycks2016gaussian,
  title={Gaussian error linear units (gelus)},
  author={Hendrycks, Dan and Gimpel, Kevin},
  journal={arXiv preprint arXiv:1606.08415},
  year={2016}
}

@inproceedings{
liu2025rfwave,
title={{RFW}ave: Multi-band Rectified Flow for Audio Waveform Reconstruction},
author={Peng Liu and Dongyang Dai and Zhiyong Wu},
booktitle={The Thirteenth International Conference on Learning Representations},
year={2025},
}

@misc{MUSDB18HQ,
  author       = {Rafii, Zafar and
                  Liutkus, Antoine and
                  Fabian-Robert St{\"o}ter and
                  Mimilakis, Stylianos Ioannis and
                  Bittner, Rachel},
  title        = {{MUSDB18-HQ} - an uncompressed version of MUSDB18},
  month        = dec,
  year         = 2019,
  doi          = {10.5281/zenodo.3338373},
}

@inproceedings{
lee2025periodwave,
title={PeriodWave: Multi-Period Flow Matching for High-Fidelity Waveform Generation},
author={Sang-Hoon Lee and Ha-Yeong Choi and Seong-Whan Lee},
booktitle={The Thirteenth International Conference on Learning Representations},
year={2025},
}

@INPROCEEDINGS{9763903,
  author={Yang, Ximei and Guendel, Ronny G. and Yarovoy, Alexander and Fioranelli, Francesco},
  booktitle={2022 IEEE Radar Conference (RadarConf22)}, 
  title={Radar-based Human Activities Classification with Complex-valued Neural Networks}, 
  year={2022},
  volume={},
  number={},
  pages={1-6},
  doi={10.1109/RadarConf2248738.2022.9763903}}

@ARTICLE{9766131,
  author={Xu, Jie and Wu, Chengyu and Ying, Shuangshuang and Li, Hui},
  journal={IEEE Access}, 
  title={The Performance Analysis of Complex-Valued Neural Network in Radio Signal Recognition}, 
  year={2022},
  volume={10},
  number={},
  pages={48708-48718},
  doi={10.1109/ACCESS.2022.3171856}}

@ARTICLE{10637717,
  author={Nustede, Eike J. and Anemüller, Jörn},
  journal={IEEE/ACM Transactions on Audio, Speech, and Language Processing}, 
  title={On the Generalization Ability of Complex-Valued Variational U-Networks for Single-Channel Speech Enhancement}, 
  year={2024},
  volume={32},
  number={},
  pages={3838-3849},
  doi={10.1109/TASLP.2024.3444492}}

@inproceedings{mamun23_interspeech,
  title     = {CFTNet: Complex-valued Frequency Transformation Network for Speech Enhancement},
  author    = {Nursadul Mamun and John H. L. Hansen},
  year      = {2023},
  booktitle = {Interspeech 2023},
  pages     = {809--813},
  doi       = {10.21437/Interspeech.2023-280},
  issn      = {2958-1796},
}

@article{Wirtinger1927,
  author    = {Wirtinger, W.},
  title     = {Zur formalen Theorie der Funktionen von mehr komplexen Veränderlichen},
  journal   = {Mathematische Annalen},
  volume    = {97},
  number    = {1},
  pages     = {357--375},
  year      = {1927},
  month     = {December},
  issn      = {1432-1807},
  doi       = {10.1007/BF01447872},
}

@inproceedings{7472755,
  author={Hu, Qiong and Yamagishi, Junichi and Richmond, Korin and Subramanian, Kartick and Stylianou, Yannis},
  booktitle={2016 IEEE International Conference on Acoustics, Speech and Signal Processing (ICASSP)}, 
  title={Initial investigation of speech synthesis based on complex-valued neural networks}, 
  year={2016},
  volume={},
  number={},
  pages={5630-5634},
  doi={10.1109/ICASSP.2016.7472755}}

@article{bengio2013estimating,
  title={Estimating or propagating gradients through stochastic neurons},
  author={Bengio, Yoshua},
  journal={arXiv preprint arXiv:1305.2982},
  year={2013}
}

@inproceedings{NEURIPS2020_11604531,
 author = {Ziyin, Liu and Hartwig, Tilman and Ueda, Masahito},
 booktitle = {Advances in Neural Information Processing Systems},
 editor = {H. Larochelle and M. Ranzato and R. Hadsell and M.F. Balcan and H. Lin},
 pages = {1583--1594},
 publisher = {Curran Associates, Inc.},
 title = {Neural Networks Fail to Learn Periodic Functions and How to Fix It},
 volume = {33},
 year = {2020}
}

@inproceedings{steinmetz2020auraloss,
  title={auraloss: Audio focused loss functions in PyTorch},
  author={Steinmetz, Christian J and Reiss, Joshua D},
  booktitle={Digital music research network one-day workshop (DMRN+ 15)},
  year={2020}
}

@inproceedings{8659610,
  author={Hayakawa, Daichi and Masuko, Takashi and Fujimura, Hiroshi},
  booktitle={2018 Asia-Pacific Signal and Information Processing Association Annual Summit and Conference (APSIPA ASC)}, 
  title={Applying Complex-Valued Neural Networks to Acoustic Modeling for Speech Recognition}, 
  year={2018},
  volume={},
  number={},
  pages={1725-1731},
  doi={10.23919/APSIPA.2018.8659610}}

@article{yoneyama2024wavehax,
  title={Wavehax: Aliasing-Free Neural Waveform Synthesis Based on 2D Convolution and Harmonic Prior for Reliable Complex Spectrogram Estimation},
  author={Yoneyama, Reo and Miyashita, Atsushi and Yamamoto, Ryuichi and Toda, Tomoki},
  journal={arXiv preprint arXiv:2411.06807},
  year={2024}
}

@INPROCEEDINGS{10448291,
  author={Mehta, Shivam and Tu, Ruibo and Beskow, Jonas and Székely, Éva and Henter, Gustav Eje},
  booktitle={ICASSP 2024 - 2024 IEEE International Conference on Acoustics, Speech and Signal Processing (ICASSP)}, 
  title={Matcha-TTS: A Fast TTS Architecture with Conditional Flow Matching}, 
  year={2024},
  volume={},
  number={},
  pages={11341-11345},
  doi={10.1109/ICASSP48485.2024.10448291}}

@InProceedings{Selvaraju_2017_ICCV,
author = {Selvaraju, Ramprasaath R. and Cogswell, Michael and Das, Abhishek and Vedantam, Ramakrishna and Parikh, Devi and Batra, Dhruv},
title = {Grad-CAM: Visual Explanations From Deep Networks via Gradient-Based Localization},
booktitle = {Proceedings of the IEEE International Conference on Computer Vision (ICCV)},
month = {Oct},
year = {2017}
}

@INPROCEEDINGS{9413814,
  author={Barrachina, J. A. and Ren, C. and Morisseau, C. and Vieillard, G. and Ovarlez, J.-P.},
  booktitle={ICASSP 2021 - 2021 IEEE International Conference on Acoustics, Speech and Signal Processing (ICASSP)}, 
  title={Complex-Valued Vs. Real-Valued Neural Networks for Classification Perspectives: An Example on Non-Circular Data}, 
  year={2021},
  volume={},
  number={},
  pages={2990-2994},
  doi={10.1109/ICASSP39728.2021.9413814}}

@phdthesis{sarroff2018complex,
  title={Complex neural networks for audio},
  author={Sarroff, Andy M},
  year={2018},
  school={Dartmouth College}
}

@article{VOIGTLAENDER202333,
title = {The universal approximation theorem for complex-valued neural networks},
journal = {Applied and Computational Harmonic Analysis},
volume = {64},
pages = {33-61},
year = {2023},
issn = {1063-5203},
doi = {https://doi.org/10.1016/j.acha.2022.12.002},
url = {https://www.sciencedirect.com/science/article/pii/S1063520322001014},
author = {Felix Voigtlaender}
}

@inproceedings{NEURIPS2023_05b69cc4,
 author = {Geuchen, Paul and Voigtlaender, Felix},
 booktitle = {Advances in Neural Information Processing Systems},
 editor = {A. Oh and T. Naumann and A. Globerson and K. Saenko and M. Hardt and S. Levine},
 pages = {1681--1737},
 publisher = {Curran Associates, Inc.},
 title = {Optimal approximation using complex-valued neural networks},
 url = {https://proceedings.neurips.cc/paper_files/paper/2023/file/05b69cc4c8ff6e24c5de1ecd27223d37-Paper-Conference.pdf},
 volume = {36},
 year = {2023}
}

@ARTICLE{10128683,
  author={Ai, Yang and Ling, Zhen-Hua},
  journal={IEEE/ACM Transactions on Audio, Speech, and Language Processing}, 
  title={APNet: An All-Frame-Level Neural Vocoder Incorporating Direct Prediction of Amplitude and Phase Spectra}, 
  year={2023},
  volume={31},
  number={},
  pages={2145-2157},
  doi={10.1109/TASLP.2023.3277276}}

@InProceedings{10.1007/978-981-97-0601-3_6,
author="Du, Hui-Peng
and Lu, Ye-Xin
and Ai, Yang
and Ling, Zhen-Hua",
editor="Jia, Jia
and Ling, Zhenhua
and Chen, Xie
and Li, Ya
and Zhang, Zixing",
title="APNet2: High-Quality and High-Efficiency Neural Vocoder with Direct Prediction of Amplitude and Phase Spectra",
booktitle="Man-Machine Speech Communication",
year="2024",
publisher="Springer Nature Singapore",
address="Singapore",
pages="66--80",
isbn="978-981-97-0601-3"
}

@inproceedings{lv24_interspeech,
  title     = {{FreeV: Free Lunch For Vocoders Through Pseudo Inversed Mel Filter}},
  author    = {Yuanjun Lv and Hai Li and Ying Yan and Junhui Liu and Danming Xie and Lei Xie},
  year      = {2024},
  booktitle = {{Interspeech 2024}},
  pages     = {3869--3873},
  doi       = {10.21437/Interspeech.2024-2407},
  issn      = {2958-1796},
}

@INPROCEEDINGS{7072087,
  author={Yu, Eric W. M. and Chan, Cheung-Fat},
  booktitle={2002 11th European Signal Processing Conference}, 
  title={Phase modeling and quantization for low-rate harmonic+noise coding}, 
  year={2002},
  volume={},
  number={},
  pages={1-4},
  doi={}}

@ARTICLE{1214851,
  author={Doh-Suk Kim},
  journal={IEEE Transactions on Speech and Audio Processing}, 
  title={Perceptual phase quantization of speech}, 
  year={2003},
  volume={11},
  number={4},
  pages={355-364},
  doi={10.1109/TSA.2003.814409}}

@article{DOU2025108685,
title = {Enhanced phase recovery in in-line holography with self-supervised complex-valued neural networks},
journal = {Optics and Lasers in Engineering},
volume = {184},
pages = {108685},
year = {2025},
issn = {0143-8166},
doi = {https://doi.org/10.1016/j.optlaseng.2024.108685},
url = {https://www.sciencedirect.com/science/article/pii/S0143816624006638},
author = {Jiazhen Dou and Qiming An and Xiaosong Liu and Yujian Mai and Liyun Zhong and Jianglei Di and Yuwen Qin}
}
\bibliographystyle{iclr2026_conference}

\appendix
\newpage

\section{Overview of Complex-Valued Neural Networks}
This section reviews the core building blocks of CVNNs—complex convolutions, activation functions, normalization, and optimization via Wirtinger calculus \citep{Wirtinger1927}.
CVNNs extend real-valued networks by jointly modeling the real and imaginary components \citep{trabelsi2018deep}. 
By preserving cross-component structure in the complex domain, they yield more coherent representations than split-channel parameterizations.

\textbf{Complex Convolutions}:  
A CVNN performs convolutions directly in the complex domain, jointly processing the real and imaginary parts.
For an input complex feature $z = x + i y$ and a complex filter $h = a + i b$, the output $z'$ of a complex convolution is:
\begin{equation}
    z' = (x * a - y * b) + i\,(x * b + y * a),
\end{equation}
where $x,y$ are the real and imaginary components of $z$, and $a, b$ are the corresponding components of $h$. Here, $*$ denotes the convolution operation applied to each channel pair before recombining.  

\textbf{Activation Functions}: 
Complex-valued networks require activation functions that handle both magnitude and phase in a coherent way.
Let $f_{\mathrm{Re}}, f_{\mathrm{Im}}, f_{\mathrm{Mag}}: \mathbb{R}\to\mathbb{R}$ be real-valued nonlinearities.
A simple split activation applies $f_{\mathrm{Re}}$ and $f_{\mathrm{Im}}$ separately to the real and imaginary components:
\begin{equation}
    f(z) = f_{\mathrm{Re}}(x) + i\,f_{\mathrm{Im}}(y),
\end{equation}
but this approach ignores the natural coupling between magnitude and phase. 
A more phase-aware alternative applies $f_{\mathrm{Mag}}$ to the magnitude and then reattaches the original phase:
\begin{equation}
     f(z) = f_{\mathrm{Mag}}\bigl(\lvert z\rvert\bigr)\,e^{i\theta},
\end{equation}
thereby preserving all phase information while still introducing the desired nonlinearity. 
(Here $|z|$ is the magnitude and $\theta$ is the phase of $z=r e^{i\theta}$.) 

\textbf{Normalization}:    
Normalization in CVNNs accounts for the joint distribution of real and imaginary components.
A general form of complex normalization is:
\begin{equation}
    z_{\text{norm}} = \frac{z - \mu}{\sigma},
\end{equation}
where $\mu$ and $\sigma$ are the mean and standard deviation of the complex input.  
To capture correlations between the real and imaginary parts, this basic normalization is extended using the covariance matrix:
\begin{equation}
    \Sigma = 
    \begin{bmatrix}
        \sigma_{xx} & \sigma_{xy} \\
        \sigma_{yx} & \sigma_{yy}
    \end{bmatrix},
\end{equation}
where $\sigma_{xx}$ and $\sigma_{yy}$ denote the variances of the real and imaginary components, respectively, and $\sigma_{xy}=\sigma_{yx}$ represents their cross-covariance.
Using the estimated covariance, the input is normalized by centering and decorrelating:
\begin{equation}
    z_{\text{norm}} = \Sigma^{-1/2} (z - \mu),
\end{equation}
and an affine transformation is then applied to restore the network’s ability to shift and scale the normalized features:
\begin{equation}
    z' = \gamma z_{\text{norm}} + \beta,
\end{equation}
where $\gamma$ and $\beta$ are learnable complex-valued parameters. 
This formulation can be applied to various normalizations (e.g., layer or instance normalization) while preserving the complex structure.

\textbf{Gradient Optimization}:  
Gradient computation in CVNNs requires special care due to the non-holomorphic nature of most complex-valued functions.  
To handle this, CVNNs employ Wirtinger calculus \citep{Wirtinger1927}, which defines the gradient of a real-valued loss $L(z)$ with respect to a complex variable $z = x + i y$ as:
\begin{equation}
    \frac{\partial L}{\partial z} = \frac{1}{2} \left( \frac{\partial L}{\partial x} - i \frac{\partial L}{\partial y} \right), \quad
    \frac{\partial L}{\partial \bar{z}} = \frac{1}{2} \left( \frac{\partial L}{\partial x} + i \frac{\partial L}{\partial y} \right).
\end{equation}
For real-valued objectives, only the conjugate gradient $\frac{\partial L}{\partial \bar{z}}$ is used for parameter updates, which ensures descent in the loss landscape:
\begin{equation}
    z^{(t+1)} = z^{(t)} - \eta \frac{\partial L}{\partial \bar{z}},
\end{equation}
where $\eta$ is the learning rate.

\begin{table}[ht]
\centering
\caption{Architecture used for both the CVNN and RVNN generators and discriminators. 
The two networks share the same layer structure and differ only in how complex variables are represented.}
\label{tab:rvnn_cvnn_arch}
\resizebox{0.6\linewidth}{!}{%
\begin{tabular}{l|c|c}
\toprule
 Component &  CVNN &  RVNN \\
\midrule
 Input              &  $\mathbb{C}^{1}$ &  $\mathbb{R}^{2}$ \\
 Hidden dimension   &  128              &  256 \\
 Depth              &  4                &  4 \\
 Layer type         &  Complex Linear   &  Linear \\
 Activation         &  Complex LeakyReLU &  LeakyReLU \\
 Output             &  $\mathbb{C}^{1}$ &  $\mathbb{R}^{2}$ \\
\bottomrule
\end{tabular}
}
\end{table}

\section{Investigating Real and Complex Models for Complex-Domain Generation}
\label{sec:analysis_real_complex}

To investigate how real-valued and complex-valued neural networks differ in learning distributions with coupled real–imaginary structure, we conduct a minimal generative modeling experiment based on a two-dimensional target density defined in the complex plane. 
This setting removes the influence of architectural factors specific to waveform generation and isolates the effect of the underlying parameterization. 
The target distribution contains a nontrivial correlation between its real and imaginary components, providing a simple but informative test case for comparing representational behavior.

The complex-valued models operate directly in $\mathbb{C}$, receiving a one-dimensional complex latent variable and propagating it through a stack of complex linear layers with complex activations.
The real-valued models use an equivalent architecture in depth but operate entirely in $\mathbb{R}$, starting from a two-dimensional latent input and producing two real outputs that are interpreted as the real and imaginary components of a sample.
To match representational width between the two model families, the hidden dimension of the RVNN layers is doubled relative to the CVNN.
A concise summary of these architectural differences is provided in Table~\ref{tab:rvnn_cvnn_arch}.
All networks are trained using the standard GAN objective with binary cross-entropy loss and identical optimization hyperparameters.

\begin{figure*}[ht]
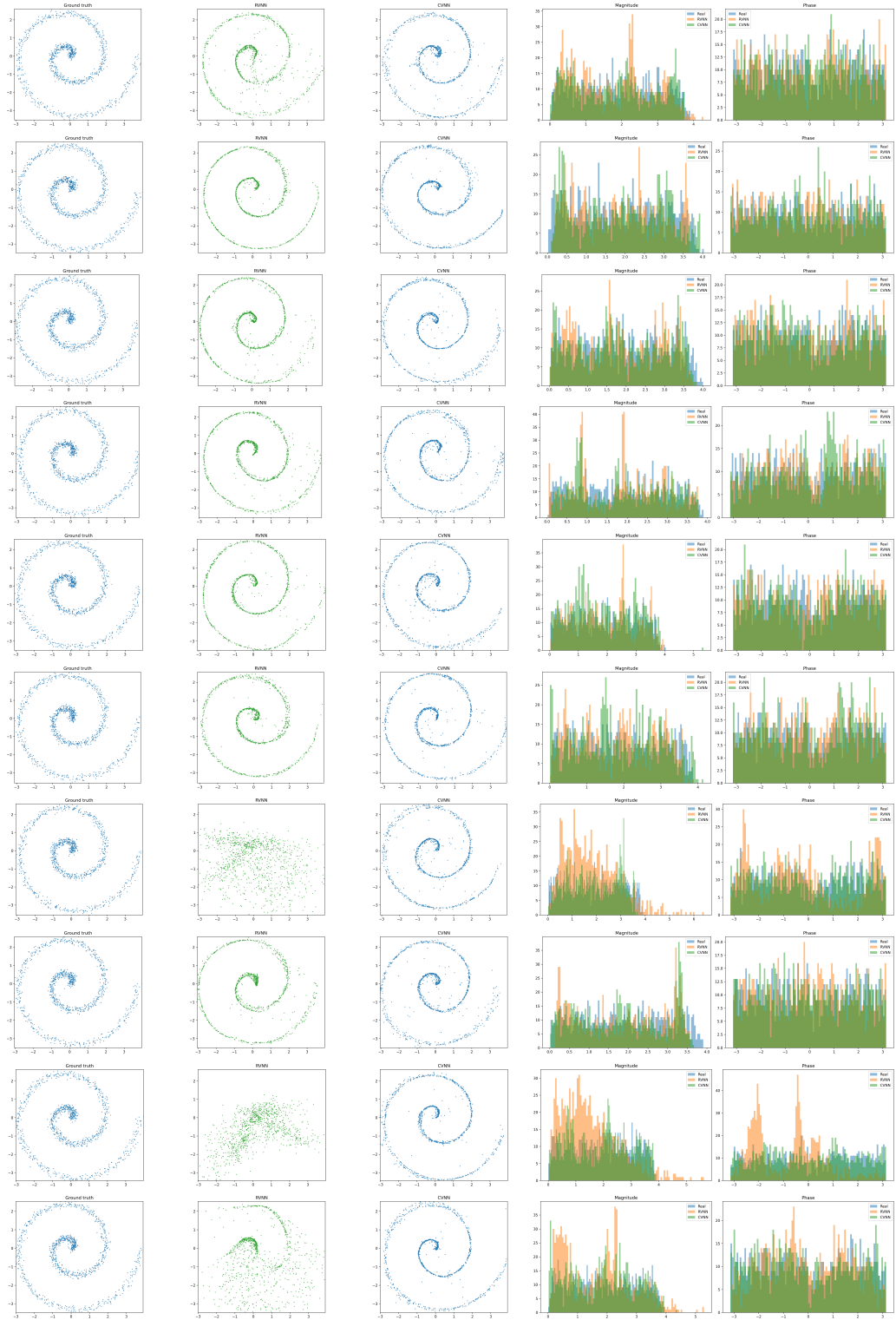

\centering
\includegraphics[width=\linewidth]{assets/mlpgan/seed0.png}

\includegraphics[width=\linewidth]{assets/mlpgan/seed1.png}

\includegraphics[width=\linewidth]{assets/mlpgan/seed2.png}

\includegraphics[width=\linewidth]{assets/mlpgan/seed3.png}

\includegraphics[width=\linewidth]{assets/mlpgan/seed4.png}

\includegraphics[width=\linewidth]{assets/mlpgan/seed5.png}

\includegraphics[width=\linewidth]{assets/mlpgan/seed6.png}

\includegraphics[width=\linewidth]{assets/mlpgan/seed7.png}

\includegraphics[width=\linewidth]{assets/mlpgan/seed8.png}

\includegraphics[width=\linewidth]{assets/mlpgan/seed9.png}
\caption{
 Visualizations over multiple training seeds. 
Each row corresponds to one run and contains five subplots: ground-truth samples, RVNN outputs, CVNN outputs, and the corresponding magnitude and phase distributions. 
This layout enables a run-to-run comparison of distributional behavior across the two models.}
\label{fig:toy_results} % \vspace{-0.6cm}
\end{figure*}

For each random seed, we examine three aspects of the learned distribution: (i) the scatter plot of generated samples, (ii) the magnitude histogram, and (iii) the phase histogram. 
These visualizations allow us to assess how consistently each model reproduces the target structure across independent runs. 
Examples are shown in Figure~\ref{fig:toy_results}. 
While both models are capable of approximating the global geometry of the target, the complex-valued generator often produces more stable spirals and magnitude–phase statistics with reduced run-to-run variability.
This experimental design does not aim to assert broad conclusions beyond this setting, but it provides a controlled example in which complex-valued parameterization can yield advantages when the modeled data are inherently expressed in the complex domain.

\clearpage

\section{Details of Training Objective}
\label{app:objective}
The ComVo training objective integrates adversarial, reconstruction, and feature‐matching losses from both the MPD and the cMRD.

\subsection{Discriminator Loss}
We use adversarial losses to push real samples above and generated samples below the decision boundary.

\textbf{MPD Loss}:
Let $D_k^{\mathrm{MPD}}$ denote the $k$-th sub-discriminator operating on raw waveforms.
For each period $P_k$, the input segment $y$ is reshaped to $(P_k,\, T/P_k)$ to expose the periodic structure.
We use a hinge loss on the real-valued outputs:
\begin{equation}
\begin{split}
\mathcal{L}_{D}^{\mathrm{MPD}}
= \sum_{k=1}^{K} \Big[
&\mathbb{E}_{y}\big( \max(0,\, 1 - D_k^{\mathrm{MPD}}(y)) \big) \\
&\quad+\;\mathbb{E}_{\hat y}\big( \max(0,\, 1 + D_k^{\mathrm{MPD}}(\hat y)) \big)
\Big],
\end{split}
\end{equation}
where $y$ and $\hat y$ are ground-truth and generated waveform segments, respectively.

\textbf{cMRD Loss}:
We apply hinge losses independently to the real and imaginary components
to retain compatibility with standard real-valued GAN objectives,
while allowing the discriminator to operate directly in the complex domain.
For any complex quantity $u$, let $[u]_R$ and $[u]_I$ denote its real and
imaginary parts, respectively (these are operators on a single complex
output, not separate networks).
With $D_k^{\mathrm{cMRD}}$ the $k$-th sub-discriminator,
\begin{equation}
\begin{split}
\mathcal{L}_{D}^{\mathrm{cMRD}}
= \sum_{k=1}^{K} \Big[
&\tfrac{1}{2}\,\mathbb{E}_{z}\big( \max(0,\, 1 - [D_k^{\mathrm{cMRD}}(z)]_R ) + \max(0,\, 1 - [D_k^{\mathrm{cMRD}}(z)]_I ) \big) \\
&\quad+\;\tfrac{1}{2}\,\mathbb{E}_{\hat z}\big( \max(0,\, 1 + [D_k^{\mathrm{cMRD}}(\hat z)]_R ) + \max(0,\, 1 + [D_k^{\mathrm{cMRD}}(\hat z)]_I ) \big)
\Big].
\end{split}
\end{equation}

\subsection{Generator Loss}
The generator objective includes reconstruction, adversarial, and feature-matching terms.

\textbf{Mel-spectrogram Loss}:
We use an $L_1$ loss on log-scaled Mel-spectrograms:
\begin{equation}
\mathcal{L}_{\mathrm{Mel}}
= \mathbb{E}\,\big\| M(y) - M(\hat y) \big\|_{1},
\end{equation}
where $y$ and $\hat y$ denote ground-truth and generated waveforms, and $M(\cdot)$ is the log-Mel transform.

\textbf{Adversarial Generator Loss}:
For the MPD operating on waveform segments $\hat y$:
\begin{equation}
\mathcal{L}_{G}^{\mathrm{MPD}}
= \sum_{k=1}^{K}
\mathbb{E}_{\hat y}\big( \max(0,\, 1 - D_k^{\mathrm{MPD}}(\hat y)) \big).
\end{equation}

For the cMRD operating on generated spectrograms $\hat z$, let $[\,\cdot\,]_R$ and $[\,\cdot\,]_I$ denote the real and imaginary parts of a complex output.
We apply hinge losses to both components:
\begin{equation}
\begin{split}
\mathcal{L}_{G}^{\mathrm{cMRD}}
= \sum_{k=1}^{K} \tfrac{1}{2}\,
\mathbb{E}_{\hat z}\Big(
&\max(0,\, 1 - [D_k^{\mathrm{cMRD}}(\hat z)]_R)
+
\max(0,\, 1 - [D_k^{\mathrm{cMRD}}(\hat z)]_I)
\Big).
\end{split}
\end{equation}

\textbf{Feature Matching Loss}:
We match intermediate representations in both discriminators.

For MPD (waveform segments $y$ and $\hat y$), we use an $\ell_1$ loss on feature maps:
\begin{equation}
\mathcal{L}_{\mathrm{FM}}^{\mathrm{MPD}}
= \sum_{k=1}^{K} \sum_{l=1}^{L_k}
\mathbb{E}\,\big\| D^{\mathrm{MPD}}_{k,l}(y) - D^{\mathrm{MPD}}_{k,l}(\hat y) \big\|_{1},
\end{equation}
where $D^{\mathrm{MPD}}_{k,l}$ is the $l$-th layer feature of the $k$-th MPD sub-discriminator.

For cMRD (complex spectrograms $z$ and $\hat z$), let $[\,\cdot\,]_R$ and $[\,\cdot\,]_I$ denote the real and imaginary parts of a complex feature, respectively.
We match the components separately:
\begin{equation}
\begin{split}
\mathcal{L}_{\mathrm{FM}}^{\mathrm{cMRD}}
= \sum_{k=1}^{K} \sum_{l=1}^{L_k} \tfrac{1}{2}\,
\mathbb{E}\Big(
&\big\| [D^{\mathrm{cMRD}}_{k,l}(z)]_R - [D^{\mathrm{cMRD}}_{k,l}(\hat z)]_R \big\|_{1} \\
&+ \big\| [D^{\mathrm{cMRD}}_{k,l}(z)]_I - [D^{\mathrm{cMRD}}_{k,l}(\hat z)]_I \big\|_{1}
\Big).
\end{split}
\end{equation}

\textbf{Total Generator Loss}:
The generator objective combines reconstruction, adversarial, and feature-matching terms:
\begin{equation}
\begin{split}
\mathcal{L}_{\mathrm{gen}}
= \;&\lambda_{\mathrm{Mel}}\,\mathcal{L}_{\mathrm{Mel}}
\;+\;\lambda_{\mathrm{MPD}}\big(\mathcal{L}_{G}^{\mathrm{MPD}}+\mathcal{L}_{\mathrm{FM}}^{\mathrm{MPD}}\big) \\
&+\;\lambda_{\mathrm{cMRD}}\big(\mathcal{L}_{G}^{\mathrm{cMRD}}+\mathcal{L}_{\mathrm{FM}}^{\mathrm{cMRD}}\big).
\end{split}
\end{equation}
Here, $\lambda_{\mathrm{Mel}}$, $\lambda_{\mathrm{MPD}}$, and $\lambda_{\mathrm{cMRD}}$ weight the Mel, MPD, and cMRD terms, respectively.
Detailed hyperparameters are provided in Table~\ref{hyperparameters}.

\section{Proof of Equivalence Between the Block-matrix Computation Scheme and Standard Complex-valued Operations}

We now verify in detail that applying the block-matrix operator
\[
A = \begin{bmatrix}
  W_r & -W_i \\[0.3em]
  W_i &  W_r
\end{bmatrix}
\]
to the stacked real vector \(\bigl[x;\,y\bigr]\) reproduces exactly the real and imaginary components of the complex product \(z' = Wz\) with \(W = W_r + i\,W_i\).

\medskip
\subsection{Forward computation}

Let
\[
z = x + i\,y,\quad W = W_r + i\,W_i,
\]
where \(x,y,W_r,W_i\) are real-valued.
Then the complex linear transformation can be written as
\begin{align*}
W\,z 
&= (W_r + iW_i)(x + i\,y) \\
&= W_r x + i\,W_i x + i\,W_r y + i^2\,W_i y \\
&= (W_r x - W_i y) \;+\; i\,(W_i x + W_r y).
\end{align*}
Thus
\[
\mathrm{Re}(z') = W_r x - W_i y,
\quad
\mathrm{Im}(z') = W_i x + W_r y.
\]
On the other hand, the block-matrix product gives
\[
A
\begin{bmatrix}x\\y\end{bmatrix}
=
\begin{bmatrix}
W_r x - W_i y \\[0.3em]
W_i x + W_r y
\end{bmatrix}
=
\begin{bmatrix}
\mathrm{Re}(z') \\[0.3em]
\mathrm{Im}(z')
\end{bmatrix}.
\]

\medskip
\subsection{Backward computation}

Let the scalar loss be \(L\), and denote
\[
g_r = \frac{\partial L}{\partial \mathrm{Re}(z')},
\quad
g_i = \frac{\partial L}{\partial \mathrm{Im}(z')}.
\]
In the complex formulation, the gradient with respect to \ \(z\) is
\[
\frac{\partial L}{\partial z}
= W^H (g_r + i\,g_i)
= \bigl(W_r^\top g_r + W_i^\top g_i\bigr)
  + i\,\bigl(-W_i^\top g_r + W_r^\top g_i\bigr).
\]
Define
\[
g_x = W_r^\top g_r + W_i^\top g_i,
\quad
g_y = -\,W_i^\top g_r + W_r^\top g_i.
\]
Stacking these gives
\begin{align}
\begin{bmatrix}
g_x \\[0.5em]
g_y
\end{bmatrix}
&=
A^\top
\begin{bmatrix}
g_r \\[0.3em]
g_i
\end{bmatrix}
=
\begin{bmatrix}
W_r & -W_i \\[0.3em]
W_i &  W_r
\end{bmatrix}^\top
\begin{bmatrix}
g_r \\[0.3em]
g_i
\end{bmatrix},
\end{align}
which is precisely the transpose of the forward block-matrix. 
For convolutional layers, each transpose block corresponds to the appropriate transposed-convolution operator.

\begin{table}[ht]
\centering
\caption{Average GPU execution times for generator (Gen) and discriminator (Disc) forward and backward passes.}
\resizebox{\textwidth}{!}{
\begin{tabular}{lrrrr}
\toprule
Implementation     & Gen Forward (s) & Gen Backward (s) & Disc Forward (s) & Disc Backward (s) \\
\midrule
Native PyTorch      & 0.005288 & 0.234591 & 0.079073 & 0.190067 \\
 Gaussian trick      &  0.005160 &  0.231545 &  0.074103 &  0.184894 \\
Block‐matrix        & 0.005786 & 0.181283 & 0.050389 & 0.139159 \\
\bottomrule
\end{tabular}
}
\label{tab:cvnn_speed}
% \vspace{-0.6cm}
\end{table}

\section{Speed Comparison of Generator and Discriminator Operations}
To isolate the effect of block-matrix fusion, we benchmark only the generator and the cMRD, excluding the MPD and reusing the same pretrained hyperparameters across all implementations.

In addition to the native PyTorch implementation and our block-matrix formulation, we also evaluated Gauss’ multiplication trick, implemented using the \texttt{complextorch} library\footnote{\url{https://github.com/josiahwsmith10/complextorch}}. 

Gauss’ multiplication trick rewrites a complex product using three real-valued convolutions instead of four, and is a common arithmetic reduction technique for complex operations.

Table~\ref{tab:cvnn_speed} reports the average GPU execution times for the forward and backward passes of both the generator and the cMRD over 10 runs with a batch size of 16.
For the generator, the forward time shows minimal variation across implementations, indicating that fusing real and imaginary components introduces little overhead in this part of the computation.
In contrast, the block-matrix formulation substantially reduces the generator’s backward time and provides clear improvements in both the forward and backward passes of the cMRD, leading to a noticeably faster end-to-end training step.
Overall, these results indicate that the block-matrix formulation can provide practical efficiency gains in our training setup.

\begin{table}[ht]
\caption{Component-level differences in intermediate values and parameter gradients between native and refined implementations.}
\label{diff_conv_linear_mean_only}
\centering
\begin{tabular}{l|c|c|c}
\toprule
Metric & Conv1d & Conv2d & Linear \\
\midrule
Input gradient              & 2e-09 & 6e-09 & 8e-10 \\
Forward output              & 4e-09 & 1e-08 & 7e-09 \\
Forward output gradient     & 9e-09 & 2e-08 & 1e-08 \\
Weight                      & 0e+00 & 0e+00 & 0e+00 \\
Weight gradient             & 1e-07 & 4e-07 & 4e-08 \\
Bias                        & 0e+00 & 0e+00 & 0e+00 \\
Bias gradient               & 6e-08 & 4e-07 & 3e-08 \\
\bottomrule
\end{tabular}
\end{table}

\begin{table}[ht]
\caption{Model-level differences in outputs, losses, and gradient magnitudes between native and refined implementations.}
\centering
\label{diff_gan_mean_only}
\begin{tabular}{l|c|c}
\toprule
Metric & Generator & Discriminator \\
\midrule
Forward output  & 7e-06 & 5e-06 \\
Loss            & 5e-07 & 2e-07 \\
Gradient        & 5e-06 & 1e-06 \\
\bottomrule
\end{tabular}
\end{table}

\section{Numerical Consistency Verification}
\label{app:numeric_consistency}
To confirm that our block-matrix computation scheme maintains numerical fidelity, we compare forward outputs and gradients for each module against the native PyTorch implementation.  
Table~\ref{diff_conv_linear_mean_only} reports mean absolute differences at the layer level for convolutional and linear modules—all within typical floating-point tolerances ($\sim10^{-7}$).  
Table~\ref{diff_gan_mean_only} summarizes end-to-end deviations in generator and discriminator outputs, losses, and gradient norms, all below $10^{-5}$.  
These results verify that, despite the structural optimizations, our block-matrix approach preserves numerical consistency and does not affect training dynamics.  

\section{Backward Graph Visualization}
\label{app:graph_vis}

\begin{figure}[!t]
\centering
\includegraphics[height=0.75\textheight, keepaspectratio]{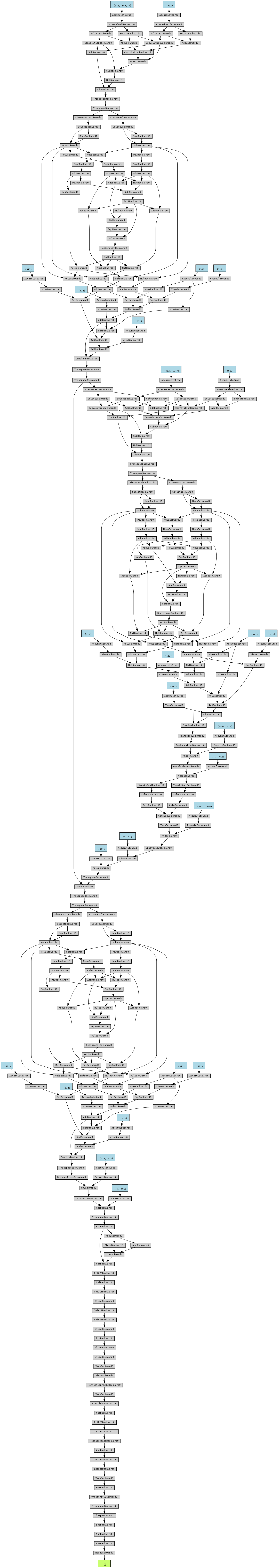}
\caption{Backward computation graph of the generator using the native PyTorch complex implementation.}
\label{fig:graph_gen_native}
\end{figure}

\begin{figure}[!t]
\centering
\includegraphics[height=0.75\textheight, keepaspectratio]{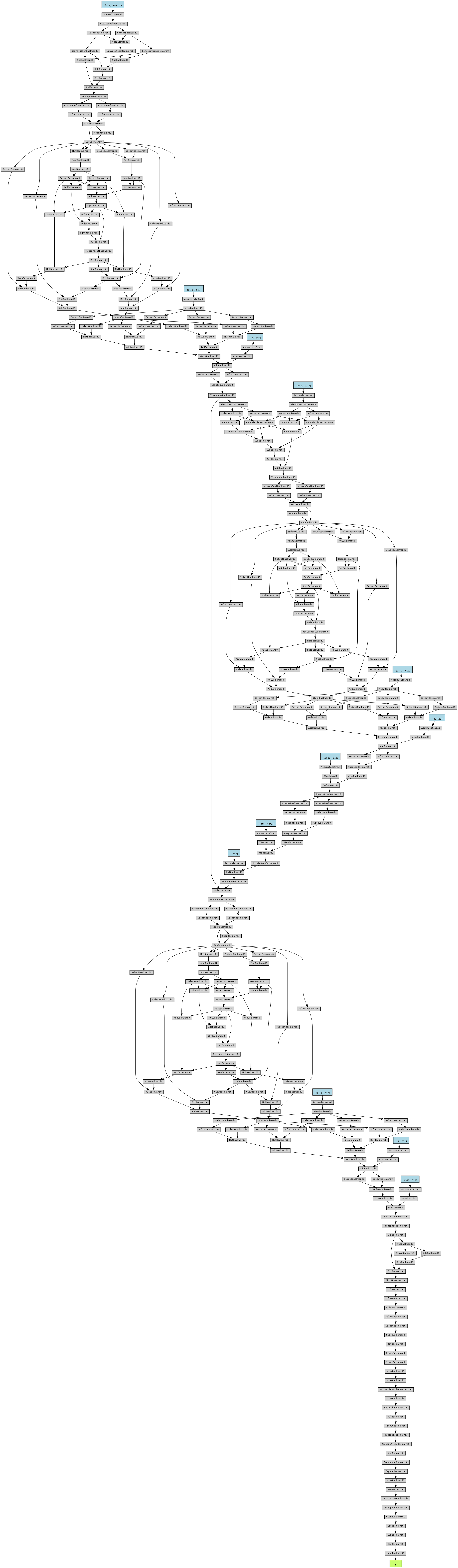}
\caption{ Backward computation graph of the generator using Gauss’ multiplication trick.}
\label{fig:graph_gen_gauss}
\end{figure}

\begin{figure}[!t]
\centering
\includegraphics[height=0.75\textheight, keepaspectratio]{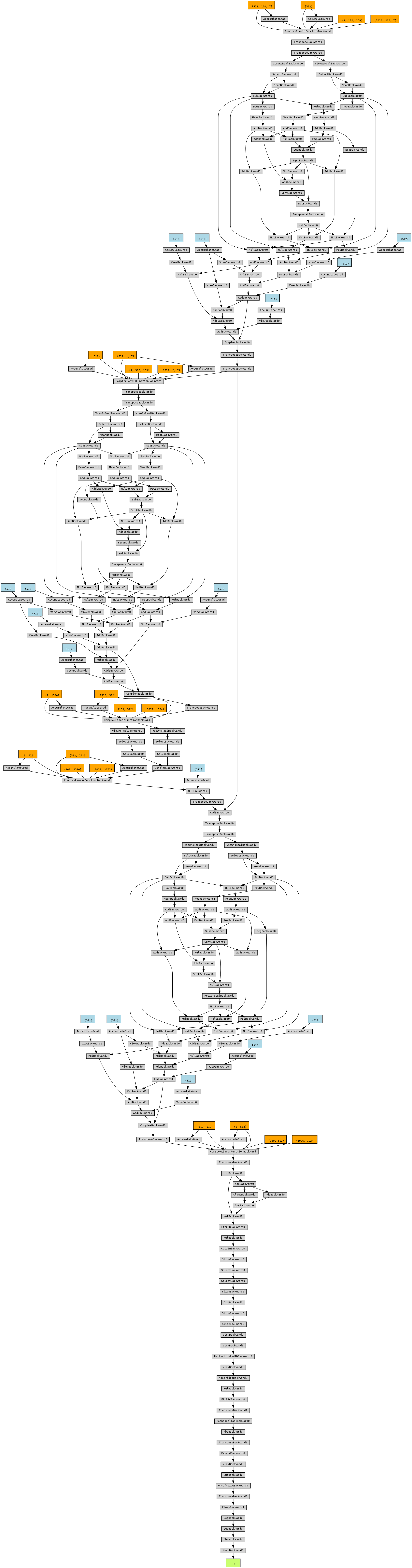}
\caption{Backward computation graph of the generator using the block-matrix operation.}
\label{fig:graph_gen_block}
\end{figure}

\begin{figure}[!t]
\centering
\includegraphics[width=0.4\textwidth, keepaspectratio]{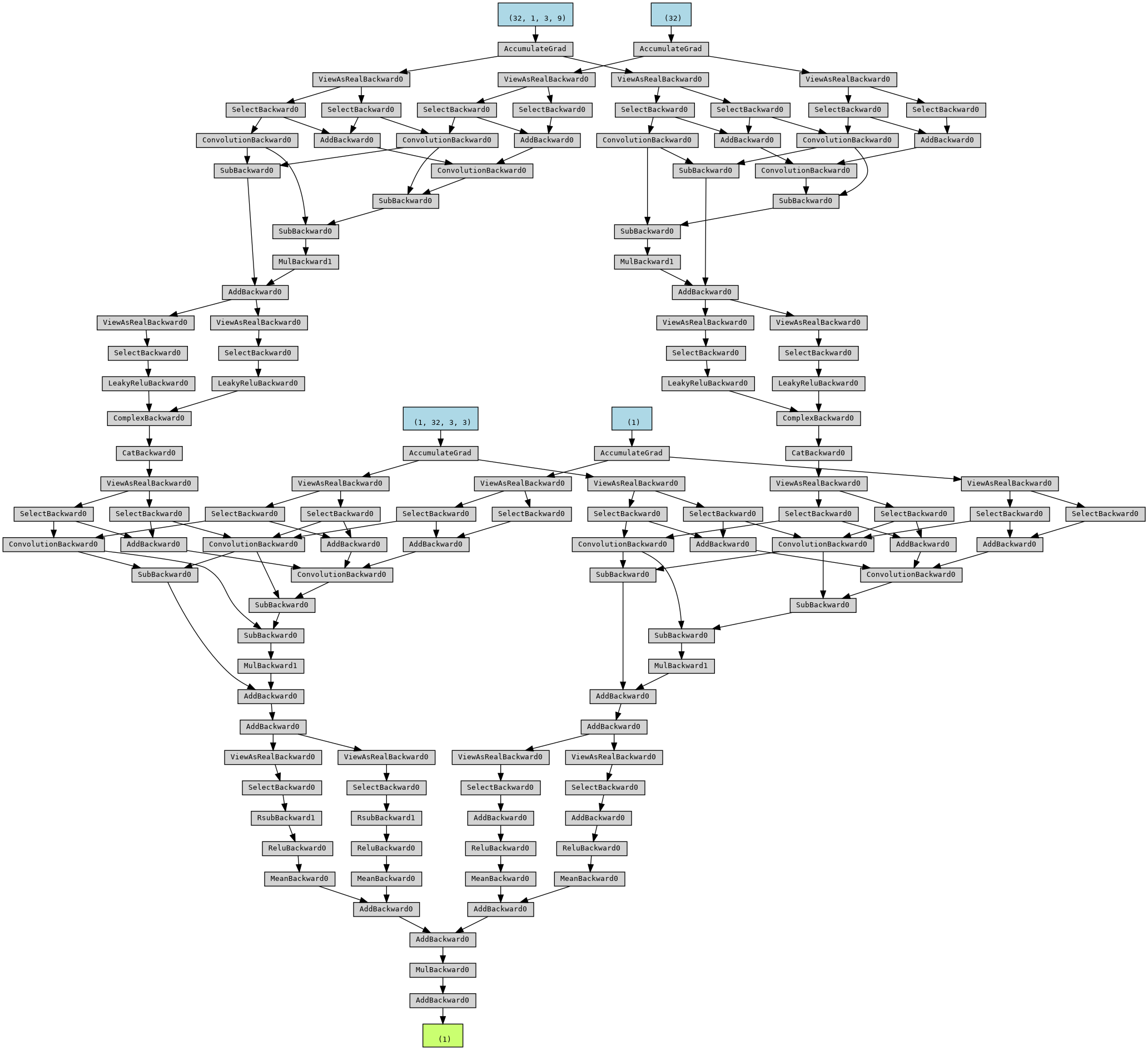}
\caption{Backward computation graph of the cMRD using the native PyTorch complex implementation.}
\label{fig:graph_d_native}
\end{figure}

\begin{figure}[!t]
\centering
\includegraphics[width=0.4\textwidth, keepaspectratio]{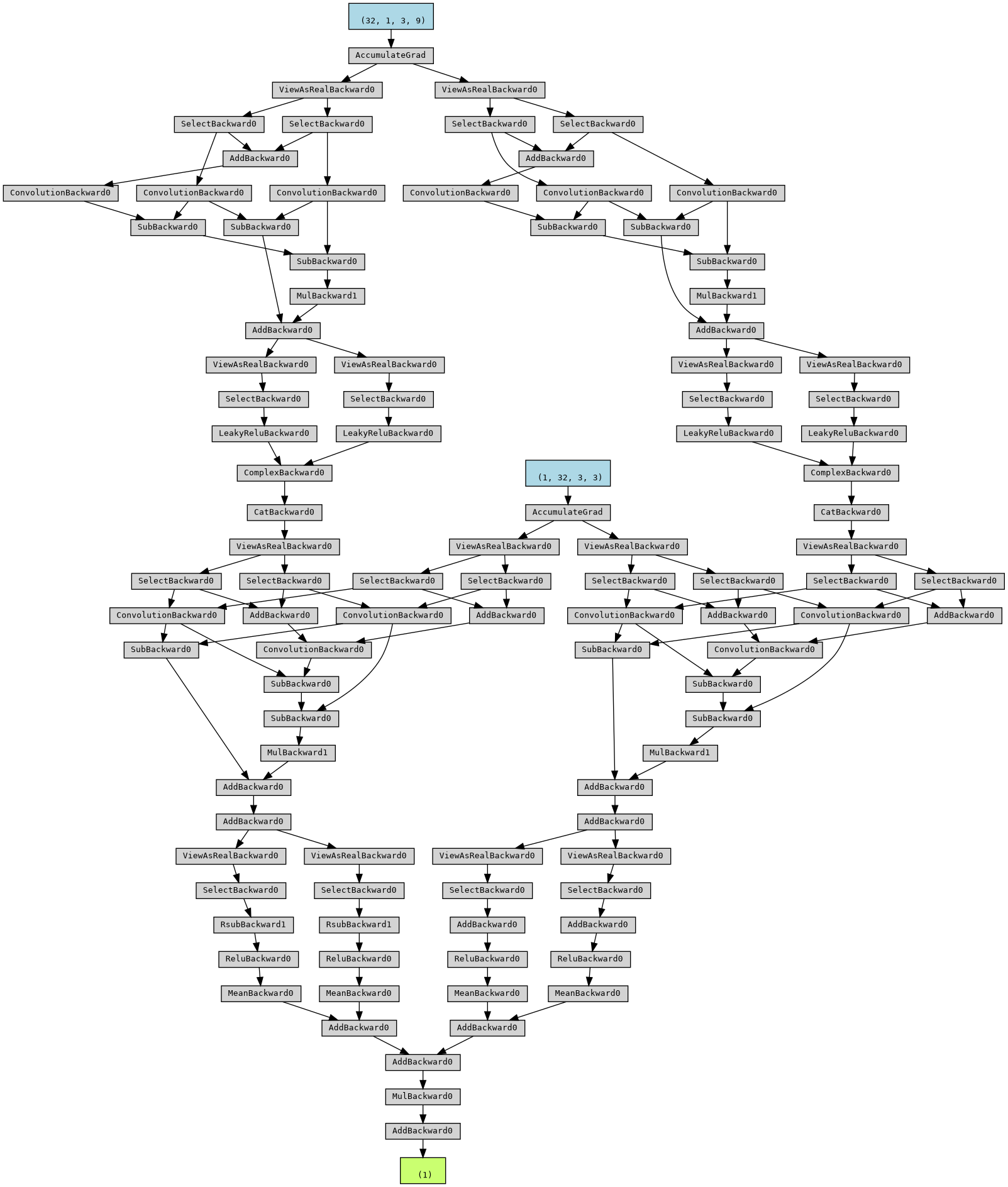}
\caption{ Backward computation graph of the cMRD using Gauss’ multiplication trick.}
\label{fig:graph_d_gauss}
\end{figure}

\begin{figure}[!t]
\centering
\includegraphics[width=0.4\textwidth, keepaspectratio]{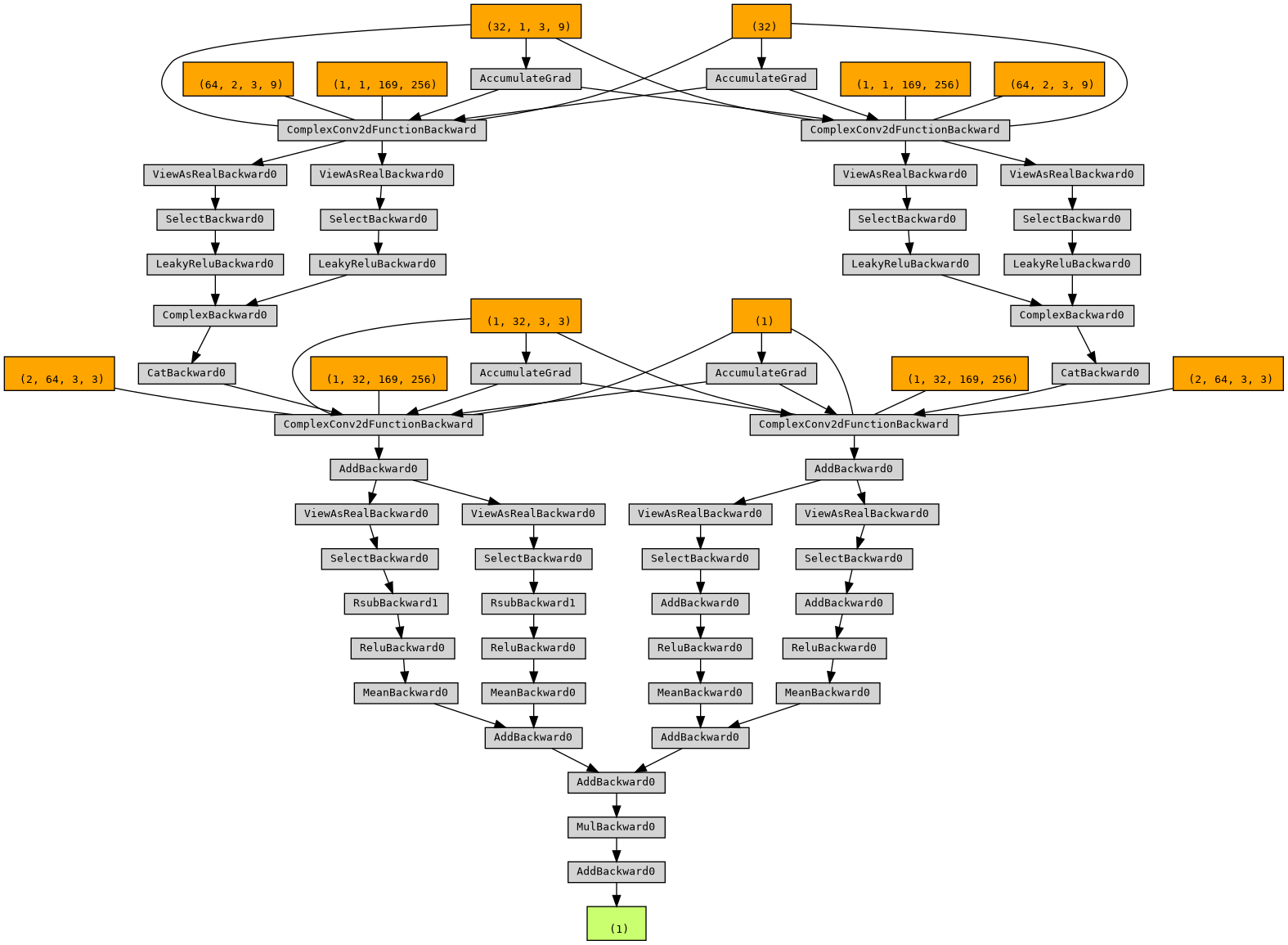}
\caption{Backward computation graph of the cMRD using the block-matrix operation.}
\label{fig:graph_d_block}
\end{figure}

Figures~\ref{fig:graph_gen_native}, \ref{fig:graph_gen_gauss}, and \ref{fig:graph_gen_block} show the backward computation graphs of the generator using
(i) the native PyTorch complex implementation,
(ii) Gauss’ multiplication trick, and
(iii) the block-matrix formulation, respectively.
Figures~\ref{fig:graph_d_native}, \ref{fig:graph_d_gauss}, and \ref{fig:graph_d_block} present the corresponding graphs for the cMRD.
For clarity, both models are simplified by using a single Mel-spectrogram loss and reducing the number of layers and channels.

Across all configurations, the block-matrix formulation (Figures~\ref{fig:graph_gen_block} and \ref{fig:graph_d_block}) yields the most compact backward graph.
Compared to the native (Figures~\ref{fig:graph_gen_native}, \ref{fig:graph_d_native}) and Gauss-based implementations (Figures~\ref{fig:graph_gen_gauss}, \ref{fig:graph_d_gauss}), it avoids redundant branches and reduces the number of elementwise operations, resulting in a significantly simpler and more efficient gradient flow.

\begin{figure*}[ht]
\centering
\includegraphics[width=0.9\linewidth]{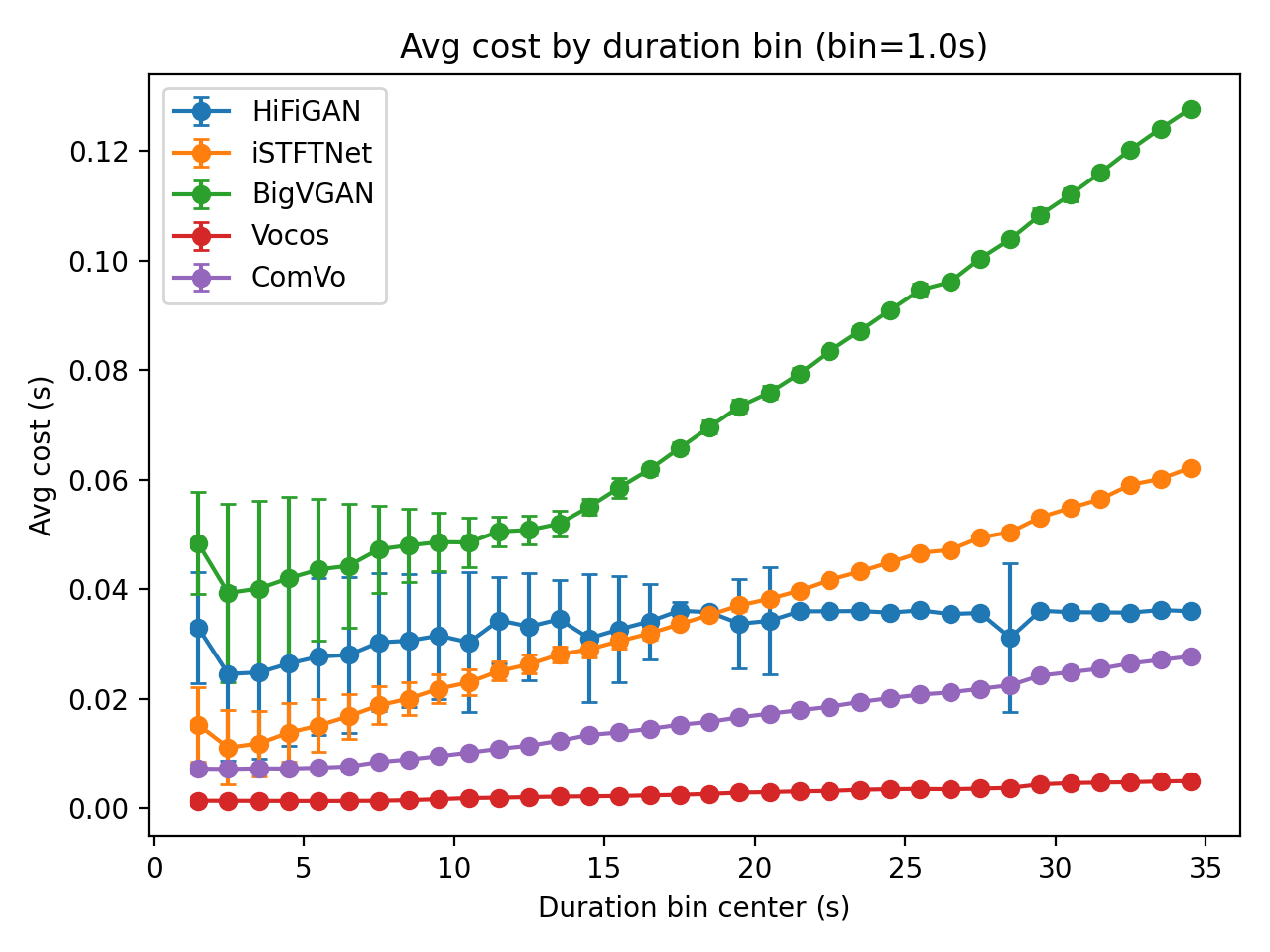}
\caption{Average inference cost as a function of utterance duration.}
\label{fig:avg_cost} % \vspace{-0.6cm}
\end{figure*}

\section{Runtime as a Function of Utterance Length}
Figure~\ref{fig:avg_cost} plots average inference cost versus utterance duration using 1-second bins and consistency under the same setup as Table~\ref{tab:computational_cost}; points indicate bin means and vertical bars show variability.
Upsampling-based vocoders increase approximately in proportion to duration with a clear positive slope, whereas iSTFT-based vocoders exhibit a flatter, near-constant profile over the plotted range.
The proposed method follows the iSTFT family: its curve lies above Vocos but remains below iSTFTNet and the upsampling-based systems across bins.
Although CVNNs introduce computational overhead, ComVo maintains competitive runtime characteristics within the iSTFT class.

\begin{table}[ht]
\centering

\caption{ Ablation comparing real Conv1D producing a two-channel (Re, Im) feature, Complex Conv w/o PQ, and Complex Conv w/ PQ}
\label{tab:pq_firstlayer}
\resizebox{0.80\linewidth}{!}{
\begin{tabular}{l|ccccc}
\toprule
Model & UTMOS $\uparrow$ & MR-STFT $\downarrow$ & PESQ $\uparrow$ & Periodicity $\downarrow$ & V/UV F1 $\uparrow$ \\
\midrule
Real Conv            & 3.6337 & 0.8610 & 3.7774 & 0.0980 & 0.9574 \\
Complex Conv w/o PQ  & 3.6646 & 0.8435 & 3.7756 & 0.0915 & 0.9625 \\
Complex Conv w/ PQ   & \textbf{3.6901} & \textbf{0.8439} & \textbf{3.8239} & \textbf{0.0903} & \textbf{0.9609} \\
\bottomrule
\end{tabular}
}
\end{table}

% \begin{table}[t]
% \centering
% 
% \caption{ Effect of inserting PQ at four locations of the ConvNeXt block: 
% [1] depthwise conv output, [2] first pointwise conv output, 
% [3] GELU output, and [4] final pointwise conv output.}
% \label{tab:pq_positions}
% \resizebox{0.70\linewidth}{!}{
% \begin{tabular}{l|ccccc}
% \toprule
% Position & UTMOS & MR-STFT & PESQ & Periodicity & V/UV F1 \\
% \midrule
% ComVo & 3.6901 & 0.8439 & 3.8239 & 0.0903 & 0.9609 \\
% [1] & 3.6110 & 0.8684 & 3.7347 & 0.0954 & 0.9583 \\
% [2] & 3.6171 & 0.8693 & 3.7103 & 0.0914 & 0.9610 \\
% [3] & 3.6174 & 0.8697 & 3.7372 & 0.0959 & 0.9575 \\
% [4] & 3.6643 & 0.8622 & 3.7780 & 0.0873 & 0.9649 \\
% \bottomrule
% \end{tabular}
% }
% \end{table}

\section{Analysis of Phase Quantization}
\label{sec:pq_analysis}

The generator receives real-valued inputs, and the imaginary component of the initial complex representation must therefore be synthesized internally by the network. 
At this early stage, the phase can vary freely, as there is no signal-driven constraint guiding how the initial complex feature should be formed. 
Prior work in speech coding has also observed that unconstrained phase can introduce instability or unnecessary variability during optimization \citep{7072087, 1214851}.  
Motivated by these considerations, we insert a phase quantization (PQ) step immediately after the first complex Conv1D layer to lightly regularize the formation of the initial complex features while allowing subsequent layers to operate without explicit phase constraints.

To examine whether the effect of PQ is tied to its interaction with the first complex layer, we trained a variant where the first complex Conv1D was replaced with a real Conv1D that outputs two channels, which are then interpreted as the real and imaginary components of a complex feature. 
Aside from this modification, the architecture remains unchanged. 
This variant trains properly and produces results similar to the version without PQ, whereas the original configuration with a complex Conv1D followed by PQ achieves higher scores across all metrics (Table~\ref{tab:pq_firstlayer}).  
This comparison indicates that the benefit of PQ is associated with its placement at the point where complex features first emerge.

% We additionally evaluated the effect of inserting PQ at deeper locations within the ConvNeXt block, whose structure is  

% \begin{center}
% \small
% \(\text{Depthwise Conv} \rightarrow [1] \rightarrow 
% \text{Pointwise Conv} \rightarrow [2] \rightarrow
% \text{GELU} \rightarrow [3] \rightarrow
% \text{Pointwise Conv} \rightarrow [4]\)
% \end{center}

% PQ was applied at each of the four positions, but these placements produced inconsistent outcomes and occasionally led to degraded performance (Table~\ref{tab:pq_positions}).  
% These findings suggest that PQ is most effective when applied immediately after the initial complex transformation and does not consistently provide advantages when used in later layers.

\begin{table}[ht]
\caption{ Comparison with amplitude–phase prediction vocoders}
\label{tab:apnet_freev}
% \vskip 0.15in
\centering
\resizebox{0.73\textwidth}{!}{%
\begin{tabular}{l|cccccc}
\toprule
Model & UTMOS $\uparrow$ & MR-STFT $\downarrow$ & PESQ $\uparrow$ & Periodicity $\downarrow$ & V/UV F1 $\uparrow$\\
\midrule
 APNet  &  2.4015 &  1.3375 &  2.8457 &  0.1582 &  0.9185 \\
 APNet2  &  2.7379 &  1.1582 &  2.7748 &  0.1448 &  0.9243 \\
 FreeV &  2.6971 &  1.1782 &  2.7960 &  0.1581 &  0.9105 \\
 ComVo &  3.6901 &  0.8439 &  3.8239 &  0.0903 &  0.9609 \\
\bottomrule
\end{tabular}
}
\vspace{-0.3cm}
\end{table}

\section{Comparison with Amplitude–Phase Prediction Vocoders}
In addition to comparing against GAN-based vocoders, we also consider amplitude–phase prediction methods, in which magnitude and phase are modeled separately using real-valued networks.
Representative examples include APNet \citep{10128683}, APNet2 \citep{10.1007/978-981-97-0601-3_6}, and FreeV \citep{lv24_interspeech}, all of which treat the two components as independent regression targets.
% Amplitude–phase prediction is a common approach in neural vocoding, where the model separately estimates magnitude and phase using real-valued architectures. 
% Representative examples include APNet \citep{10128683}, APNet2 \citep{10.1007/978-981-97-0601-3_6}, and FreeV \citep{lv24_interspeech}, all of which treat the two components as independent regression targets.

To position our complex-domain formulation relative to this family of methods, we trained APNet, APNet2, and FreeV with their official implementations under the same data and training settings used in our system. 
This provides a controlled comparison between explicit amplitude–phase estimation and directly modeling complex STFT coefficients.

Table~\ref{tab:apnet_freev} presents the results. 
Across all metrics, the proposed model achieves higher quality, suggesting that learning in the complex domain is an effective parameterization for iSTFT-based generation compared to treating magnitude and phase as separate prediction targets.

\begin{table}[ht]
\caption{Baseline model implementations and sources.}
\label{tab:baselines}
\centering
\resizebox{0.8\textwidth}{!}{%
\begin{tabular}{l|l}
\toprule
\textbf{Model} & \textbf{Implementation Source} \\
\midrule
iSTFTNet & \url{https://github.com/rishikksh20/iSTFTNet-pytorch} \\
HiFi-GAN & \url{https://github.com/jik876/hifi-gan} \\
BigVGAN & \url{https://github.com/NVIDIA/BigVGAN} \\
Vocos & \url{https://github.com/gemelo-ai/vocos} \\
 APNet &  \url{https://github.com/YangAi520/APNet} \\
 APNet2 &  \url{https://github.com/redmist328/APNet2} \\
 FreeV &  \url{https://github.com/BakerBunker/FreeV} \\
\bottomrule
\end{tabular}
}
\end{table}

\section{Baseline model implementations}
\label{app:baseline}
We evaluate our proposed method against several representative neural vocoders, each with distinct architectural designs:

\textbf{HiFi-GAN (v1)} \citep{NEURIPS2020_c5d73680}:  
A GAN-based vocoder that uses multiple discriminators (MPD and MRD) with a transposed convolutional generator.  
It emphasizes high-fidelity waveform generation with fast inference.

\textbf{iSTFTNet} \citep{9746713}:  
A lightweight vocoder that replaces upsampling layers with iSTFT to reduce redundant computations.  
It directly predicts complex-valued spectrograms, simplifying the overall architecture.

\textbf{BigVGAN (base)} \citep{lee2023bigvgan}:  
An improved HiFi-GAN variant that introduces the Snake function \citep{NEURIPS2020_11604531} for better modeling of periodicity and high-frequency details.  
It also adopts a scaled discriminator design, contributing to more stable GAN training and enhanced performance on challenging inputs.

\textbf{Vocos} \citep{siuzdak2024vocos}:  
An iSTFT-based vocoder built on a ConvNeXt \citep{Liu_2022_CVPR} architecture that predicts Fourier spectral coefficients for waveform reconstruction.  
It achieves high-quality synthesis with low latency.

\textbf{APNet} \citep{10128683}:  
A vocoder that separately predicts amplitude and phase spectra using independent real-valued branches. 
Phase is modeled explicitly through a parallel estimation module with anti-wrapping losses, and the waveform is reconstructed via iSTFT.

\textbf{APNet2} \citep{10.1007/978-981-97-0601-3_6}:  
An improved version of APNet that adopts a ConvNeXt v2 backbone and multi-resolution discriminators. 
It retains the separate amplitude–phase prediction design while offering higher fidelity and greater training stability.  

\textbf{FreeV} \citep{lv24_interspeech}:  
A lightweight amplitude–phase vocoder derived from APNet2 that incorporates signal-processing priors. 
It obtains an approximate amplitude spectrum via pseudo-inverse mel filtering, reducing ASP complexity while maintaining quality.

We use the official implementations provided by the authors whenever available, except for iSTFTNet, which lacks an official repository.  
For iSTFTNet, we adopt a publicly available open-source implementation instead.  
Implementation sources are summarized in Table~\ref{tab:baselines}.

\begin{table}[ht]
\caption{Implementation sources for objective evaluation metrics.}
\label{tab:metric_sources}
\centering
\resizebox{0.9\textwidth}{!}{%
\begin{tabular}{l|l}
\toprule
\textbf{Model} & \textbf{Implementation Source} \\
\midrule
UTMOS & \url{https://github.com/sarulab-speech/UTMOS22} \\
MR-STFT & \url{https://github.com/csteinmetz1/auraloss} \\
PESQ & \url{https://github.com/ludlows/PESQ} \\
Periodicity RMSE \& V/UV F1 score & \url{https://github.com/descriptinc/cargan} \\
\bottomrule
\end{tabular}
}
\end{table}

\section{Evaluation Metrics}
\label{app:metrics}
\subsection{Subjective Evaluation}
\begin{figure*}[!t]
\centering
\includegraphics[width=\linewidth]{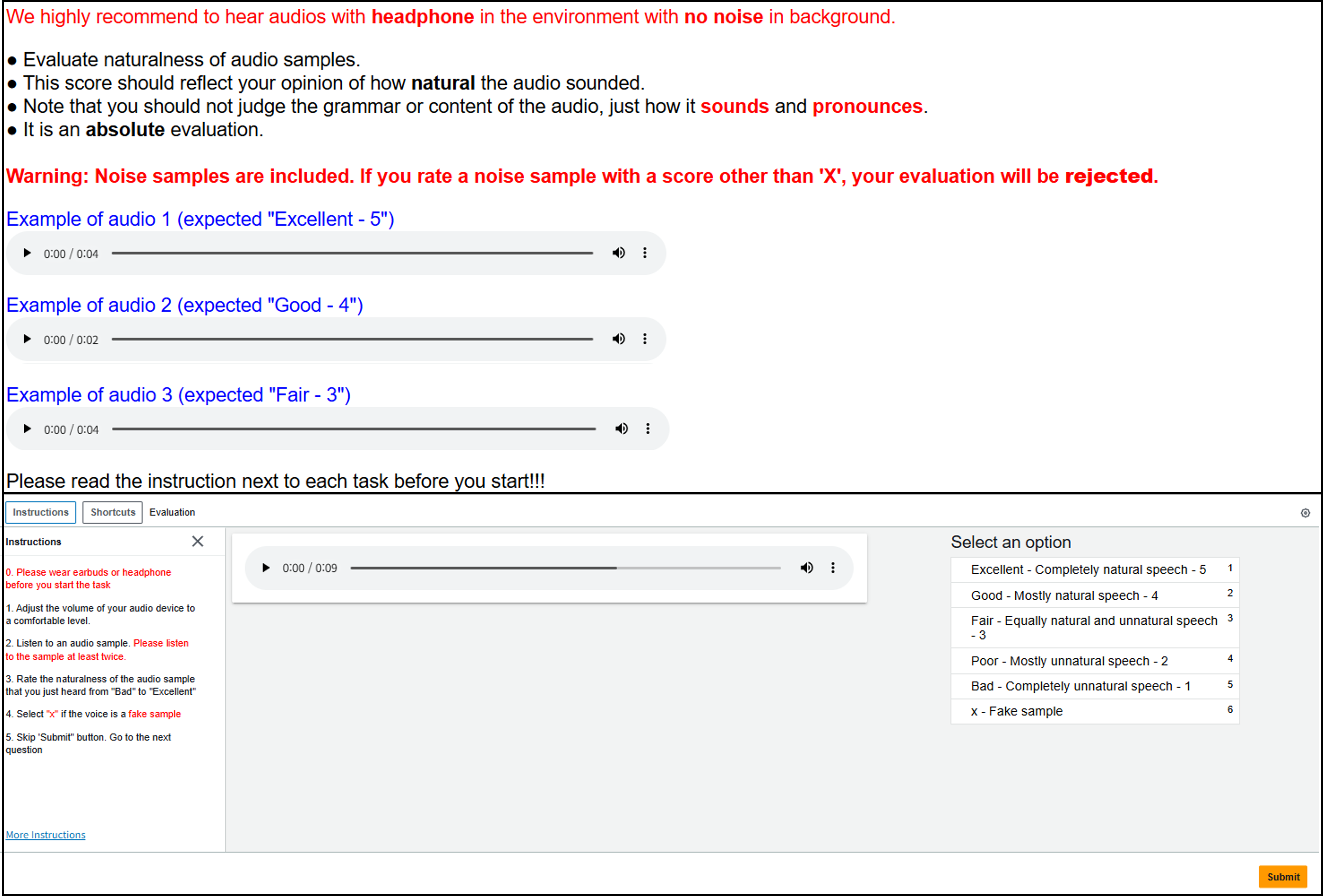}
\caption{MOS evaluation interface.}
\label{mos} % \vspace{-0.6cm}
\end{figure*}

\begin{figure*}[!t]
\centering
\includegraphics[width=\linewidth]{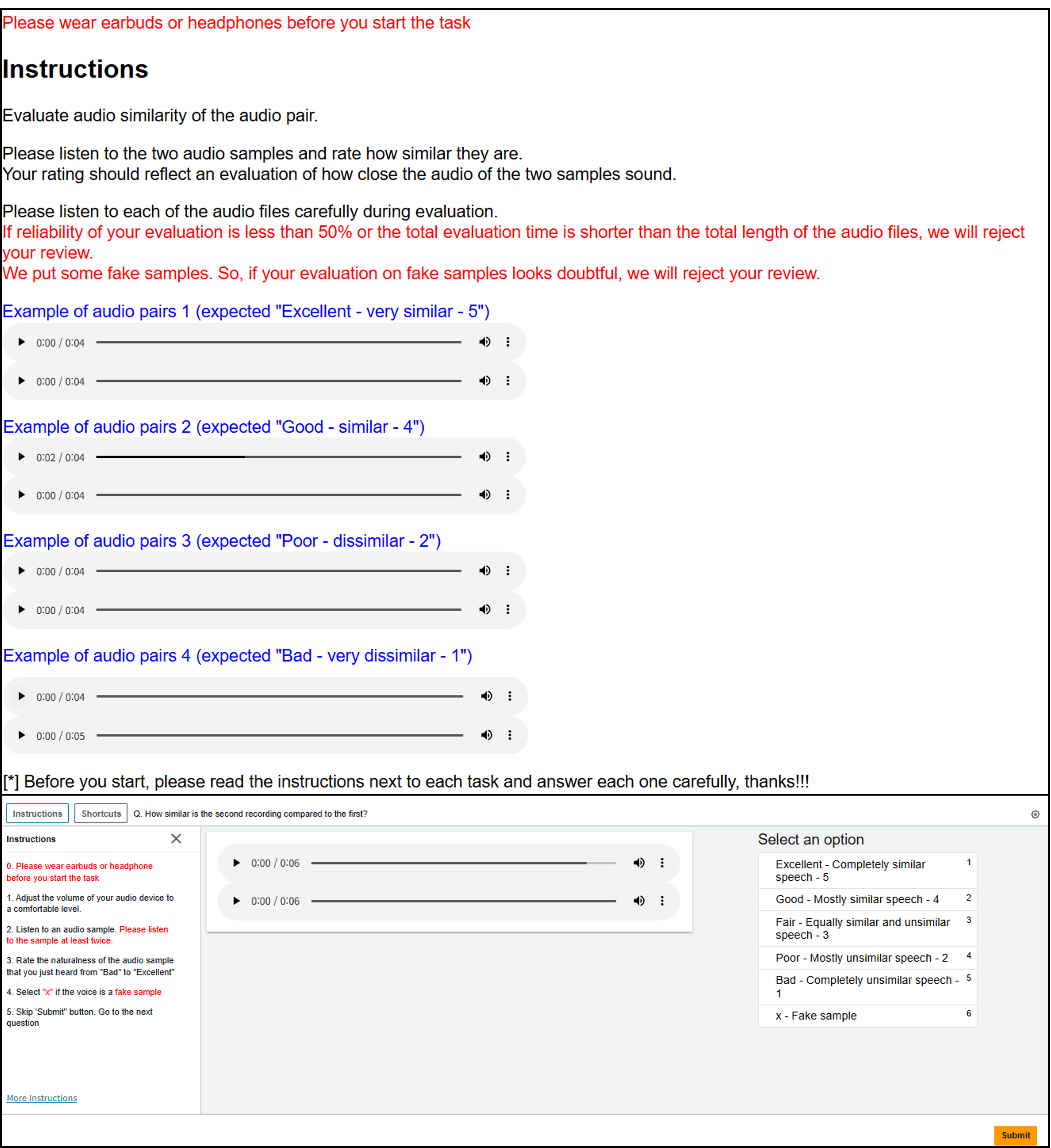}
\caption{SMOS evaluation interface.}
\label{smos} % \vspace{-0.6cm}
\end{figure*}

\begin{figure*}[!t]
\centering
\includegraphics[width=\linewidth]{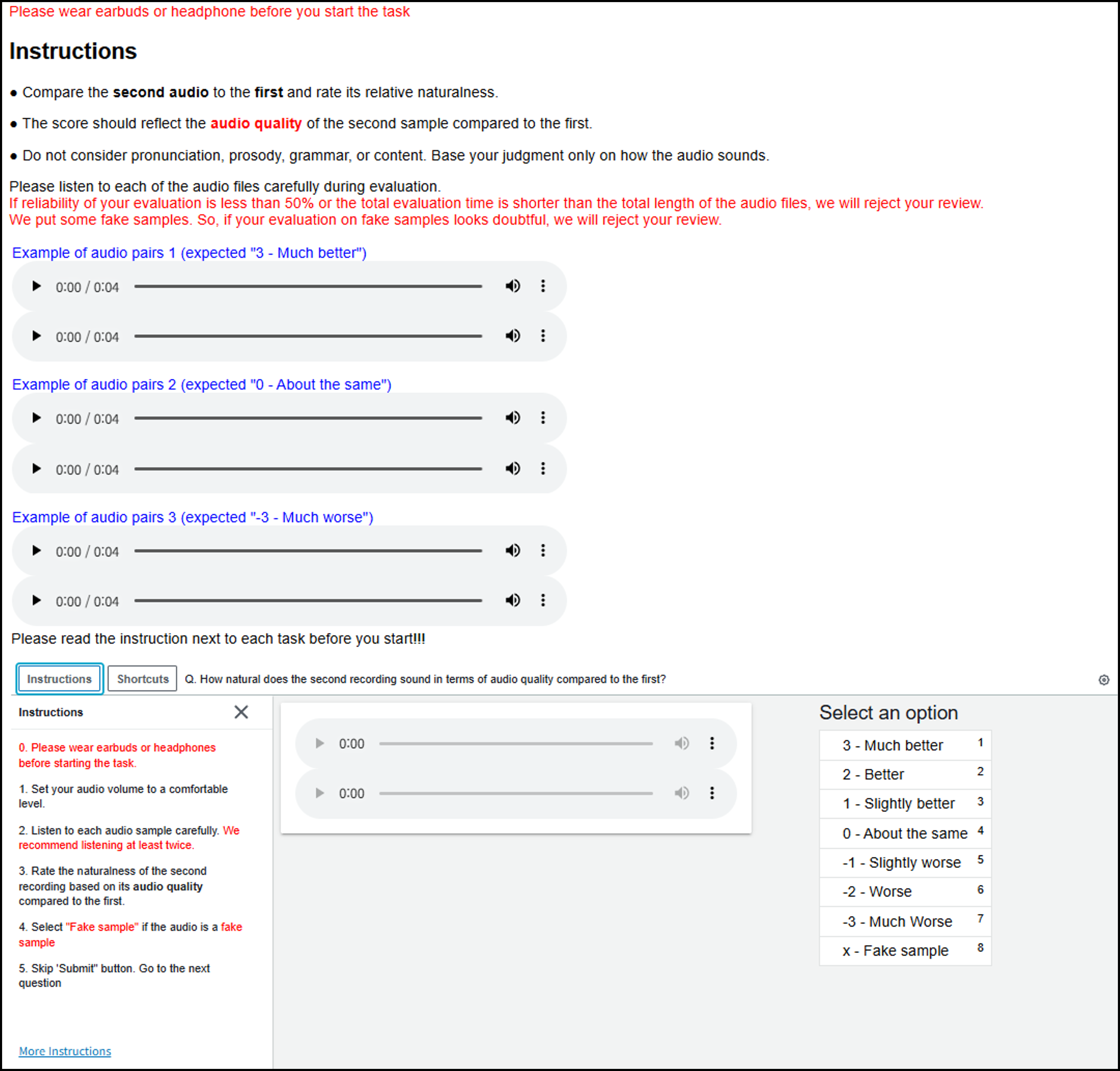}
\caption{CMOS evaluation interface.}
\label{cmos} % \vspace{-0.6cm}
\end{figure*}

We conducted mean opinion score (MOS) listening tests on Mechanical Turk with 20 U.S.-based native English speakers, each evaluating 50 samples.
We also ran similarity mean opinion score (SMOS) tests under the same conditions.
In MOS, listeners rated naturalness on a 1–5 scale; in SMOS, they rated similarity between synthesized and reference audio on a 1–5 scale.
In addition, we conducted comparison MOS (CMOS) using a 7-point scale.
For reporting, we use pairwise comparisons against our system as the reference; thus the reference row is centered at 0 and other systems’ scores reflect average preference relative to it.
To filter inattentive participants, we inserted fake samples and instructed listeners to mark them as “X”; any listener who missed these was excluded.
Figure~\ref{mos} shows the MOS interface, Figure~\ref{smos} shows the SMOS interface and Figure~\ref{cmos} shows the CMOS interface.

\subsection{Objective Evaluation}
We measure performance using five objective metrics: UTMOS \citep{saeki22c_interspeech}, multi-resolution short-time Fourier transform error (MR-STFT) \citep{9053795}, perceptual evaluation of speech quality (PESQ) \citep{941023}, periodicity RMSE, and voiced/unvoiced (V/UV) F1 score \citep{morrison2022chunked}.  
Implementation sources are listed in Table~\ref{tab:metric_sources}.

\textbf{UTMOS}: We use the open-source UTMOS model to predict MOS scores for evaluating speech naturalness.

\textbf{MR-STFT}: We use the multi-resolution STFT loss implementation from Auraloss \citep{steinmetz2020auraloss} to measure spectral distortion between the generated and ground-truth audio.  

\textbf{PESQ}: We use the wideband version of PESQ with audio resampled to 16\,kHz to assess perceptual quality.  

\textbf{Periodicity and V/UV F1}: Periodicity RMSE is used to quantify periodic artifacts, while the V/UV F1 score measures the accuracy of voiced/unvoiced classification.

\begin{table}[!h]
\caption{Comparison of large-scale models}
\label{tab:large-scale}
\centering
\resizebox{0.9\textwidth}{!}{%
\begin{tabular}{l|c|ccccc}
\toprule
Model         & Params. (M) & UTMOS $\uparrow$ & MR-STFT $\downarrow$ & PESQ $\uparrow$ & Periodicity $\downarrow$ & V/UV F1 $\uparrow$ \\
\midrule
GT              & -     & 3.8712 & -        & -   & - & -     \\
\midrule
BigVGAN (large) & 112.41 & 3.5489 & 0.8644   & 3.8197          & 0.0888       & 0.9607 \\
Vocos (large)   & 114.51 & 3.6923 & 0.8625   & 3.8362          & 0.0933       & 0.9596 \\
ComVo (large)   & 114.56  & \textbf{3.7337} & \textbf{0.8443} & \textbf{3.8831} & \textbf{0.0871}       & \textbf{0.9629} \\
\bottomrule
\end{tabular} 
} % \vspace{-0.3cm}
\end{table}

\section{Extended experiments with large-scale configurations}
To test whether the benefits of complex-valued modeling persist at higher capacity, we conducted a scaling study with large variants of the baselines and our model. 
All systems were trained on the same LibriTTS splits as in the base-scale experiments. 
For BigVGAN, we used the authors’ official large configuration; for Vocos and ComVo, we set configurations to match the BigVGAN large model’s parameter budget as closely as possible while keeping architectures comparable. 
All runs were trained for 1M optimization steps on a single GPU.
Table~\ref{tab:large-scale} summarizes the large-scale results. 
In this setting, ComVo scaled effectively, showing clear quality gains across evaluation metrics.
Overall, the complex-valued approach scales well, and increasing capacity yields consistent quality gains.

\begin{table}[!t]
\caption{Training hyperparameters.}
\label{hyperparameters}
\centering
% \resizebox{0.95\textwidth}{!}{
\begin{tabular}{lcc}
\toprule
\multicolumn{3}{l}{\textbf{Mel-spectrogram}}\\
\cmidrule(lr){1-3}
Sampling rate            & \multicolumn{2}{c}{24,000} \\
FFT size                 & \multicolumn{2}{c}{1024} \\
Hop length               & \multicolumn{2}{c}{256} \\
Window size              & \multicolumn{2}{c}{1024} \\
Mel bins                 & \multicolumn{2}{c}{100} \\
\midrule
\multicolumn{1}{l}{\textbf{Generator}} & \textbf{Base} & \textbf{Large} \\
\cmidrule(lr){1-3}
Input channels           & 100   & 100   \\
Model dimension          & 512   & 1536  \\
Intermediate dimension   & 1536  & 4608  \\
Number of layers         & 8     & 8     \\
Phase quantization levels& 128   & 128   \\
\midrule
\multicolumn{3}{l}{\textbf{MPD}}\\
\cmidrule(lr){1-3}
Periods $P_k$            & \multicolumn{2}{c}{[2, 3, 5, 7, 11]} \\
\midrule
\multicolumn{3}{l}{\textbf{MRD / cMRD}}\\
\cmidrule(lr){1-3}
FFT sizes                & \multicolumn{2}{c}{[512, 1024, 2048]} \\
Hop sizes                & \multicolumn{2}{c}{[128, 256, 512]} \\
Window sizes             & \multicolumn{2}{c}{[512, 1024, 2048]} \\
Bands ratio              & \multicolumn{2}{c}{[0, 0.1, 0.25, 0.5, 0.75, 1.0]} \\
\midrule
\multicolumn{1}{l}{\textbf{Training}} & \textbf{Base} & \textbf{Large} \\
\cmidrule(lr){1-3}
Batch size               & 16     & 32     \\
Steps                    & 1M     & 1M     \\
Segment size             & 16,384 & 16,384 \\
Initial learning rate    & 2e-4   & 2e-4   \\
Scheduler                & cosine & cosine \\
Optimizer                & AdamW  & AdamW  \\
$\beta_1, \beta_2$       & (0.8, 0.9) & (0.8, 0.9) \\
$\lambda_{\mathrm{Mel}}$ & 45     & 45     \\
$\lambda_{\mathrm{MPD}}$ & 1.0    & 1.0    \\
$\lambda_{\mathrm{cMRD}}$& 0.1    & 0.1    \\
\midrule
\multicolumn{3}{l}{\textbf{Hardware}}\\
\cmidrule(lr){1-3}
GPU                      & \multicolumn{2}{c}{1$\times$ NVIDIA A6000} \\
CPU                      & \multicolumn{2}{c}{Intel Xeon Gold 6148 @ 2.40\,GHz} \\
\bottomrule
\end{tabular}
% }
\end{table}

\newpage

\clearpage

\clearpage

\end{document}